\documentclass[journal, A4paper]{IEEEtran}
\usepackage{nicematrix, cite,graphicx,subfigure,mhchem,amsmath, mathtools}
\usepackage{color, colortbl}
\usepackage[english]{babel}
\usepackage[top=1.5cm,bottom=1.5cm,left=1.5cm,right=1.5cm,marginparwidth=1.75cm]{geometry}
\usepackage[colorlinks=true, allcolors=blue]{hyperref}

\definecolor{Gray}{gray}{0.9}

\title{In-situ biological ozone detection by measuring electrochemical impedances of plant tissues}
\author{Serge Kernbach \\[3mm]
\small CYBRES GmbH, Research Center of Advanced Robotics and Environmental Science,\\
\small Melunerstr. 40, 70569 Stuttgart, Germany, {\it serge.kernbach@cybertronica.de.com}
\vspace{-7mm}
}

\begin{document}
\maketitle

\begin{abstract}
This work demonstrates biological detection of a low concentration of \ce{O_3} by measuring electrochemical impedances of tissues in tobacco and tomato plants located indoor and outdoor. The lower range of generated ozone in the \ce{O_3}-air mix is about 30 \ce{\mu gm^{-3}} over the atmospheric level, which allows phytosensors to be considered as biodetectors of environmental pollutants. The ozone stress affects stomatal regulation that in turn influences the hydrodynamics of fluid transport system in plants. Sensors utilize electrochemical impedance spectroscopy (EIS) to measure ionic fluid content at several positions on the plant stem and calculate a variation of fluid distribution in control and experimental cases indoors. Outdoor setup uses the same methodology and sensors but different analysis due to uncontrolled nature of ozone pollution and the overlap of various stressors. The measurement results indicate a qualitative and quantitative reaction of hydrodynamic system to changes in \ce{O_3} concentration in the upper part of stem with a delay of 10-20 minutes between the onset of exposure and biological response. The probability of false-negative responses from a single plant is about 0.15$\pm$0.06. Pooling data from at least three plants allows for 92\% confidence in detecting excess \ce{O_3}. Measurements on days with low and high ozone levels of 80 \ce{\mu gm^{-3}} to 130 \ce{\mu gm^{-3}} result in a 2.33-fold difference in sensor values and thus demonstrate biological detection of high \ce{O_3} also outdoors. Statistically significant data include 948 sensor-plant attempts during 51 days with 9 plants and about 10$^7$ samples collected in automated experiments. Long-term measurements have demonstrated the high reliability of electrochemical sensors, especially in harsh outdoor conditions with rain, heat and UV/IR radiations. The described approach has applications in environmental monitoring, biological pollution detection and biosensing; low-cost EIS sensors can be utilized in precise agriculture and vertical farming to detect nonspecific biotic and abiotic stressors. 
\end{abstract}


\section{Introduction}

\ce{O_3} is one of the major environmental pollutants \cite{agronomy11081504}, \cite{Karmakar22}, it triggers various reactions in plant organisms \cite{Morales21}, \cite{DUSART20191687}, affects stomatal regulation \cite{Hasan21} and leads to different physiological changes \cite{Hoshika15}. Biological detection of ozone (as well as other environmental pollutants) can be conducted by measuring such physiological responses in-situ \cite{10.1145/3462203.3475885}, \cite{WatchPlant21}. It has already been demonstrated that high \ce{O_3} values ($>$1ppm) can be reliably detected by sensing electrophysiological parameters such as biopotentials and tissue impedances \cite{Buss23}. Focus on electrophysiological measurements is explained by their practical relevance, among others, the simplicity and reliability of sensors, enabling outdoor and field applications.   

This work extends this approach for lower \ce{O_3} concentrations, following variations of surface atmospheric \ce{O_3} at 40-50 ppbv level \cite{Kerr20}, \cite{atmos15070852}. Since changes in stomata conductance induced by \ce{O_3} in turn alter hydrodynamic parameters of water-sap transport in the plant stem, we apply electrochemical impedance spectroscopy (EIS) \cite{kernbach2022electrochemical} to capture fluid dynamics in stem tissues. EIS has a number of applications in plant science \cite{Liu21}, \cite{Jin15}, allowing to record various physiological parameters \cite{Haeverbeke23}. In this work, differential EIS is used for two-point measurements -- in the lower and upper regions of stem -- representing different parameters of the fluid transportation system \cite{kernbach2024Biohybrid}. In addition, leaf transpiration and several environmental parameters are recorded.

For biosensing, we use tomato and tobacco plants, motivated by previous research and a high sensitivity of tobacco to \ce{O_3} stress \cite{Agathokleous23}, \cite{Heggestad66}. Tobacco has already been employed as a biosensor for \ce{O_3} monitoring \cite{Calatayud07}. The generation and sensing of low \ce{O_3} concentrations represents several technological issues. We utilize four \ce{O_3} sensors, high resolution data loggers, calibrations and a PWM-controlled \ce{O_3} generator to produce \ce{O_3}-air mix with the required ozone concentration.

In addition, long-term measurements were performed under outdoor conditions with tomato plants during summer with high (120-149 \ce{\mu gm^{-3}}) and low (60-80 \ce{\mu gm^{-3}}) concentrations of atmospheric \ce{O_3}. Outdoor setup is characterized by overlapping of several stressors (\ce{O_3}, heat, NOx, PMx, UV/IR, mechanical impact) affecting plants from physiological level up to the level of physical chemistry \cite{WANG2021100397}. The indoor setup allows isolating only \ce{O_3} stress, exploring different types of physiological reactions and performing a sufficient number of iterative experiments to collect statistically significant data.

All experiments were conducted automatically with continuous recording of all biological and environmental parameters and pre-programmed timing/PWM settings of generated \ce{O_3}-air mix. Human intervention in both setups during 51 days of experiments was minimal. With two exposures per day and 6 plants, we accumulated data from about 768 indoor plant-sensor trials. The outdoor setup with 3 plants produced 180 measurement attempts. The described approach has several applications for environmental monitoring such as biological pollution detection, biosensing and biohybrid system development. Furthermore, low cost EIS sensors can be utilized in precise agriculture and vertical farming \cite{kernbach2024Biohybrid} for detecting different biotic and abiotic stressors or metabolic activities \cite{Kernbach18Yeasten}.

\section{Methods and setup}

\begin{figure*}[htp]
\centering
\subfigure{\includegraphics[width=1.0\textwidth]{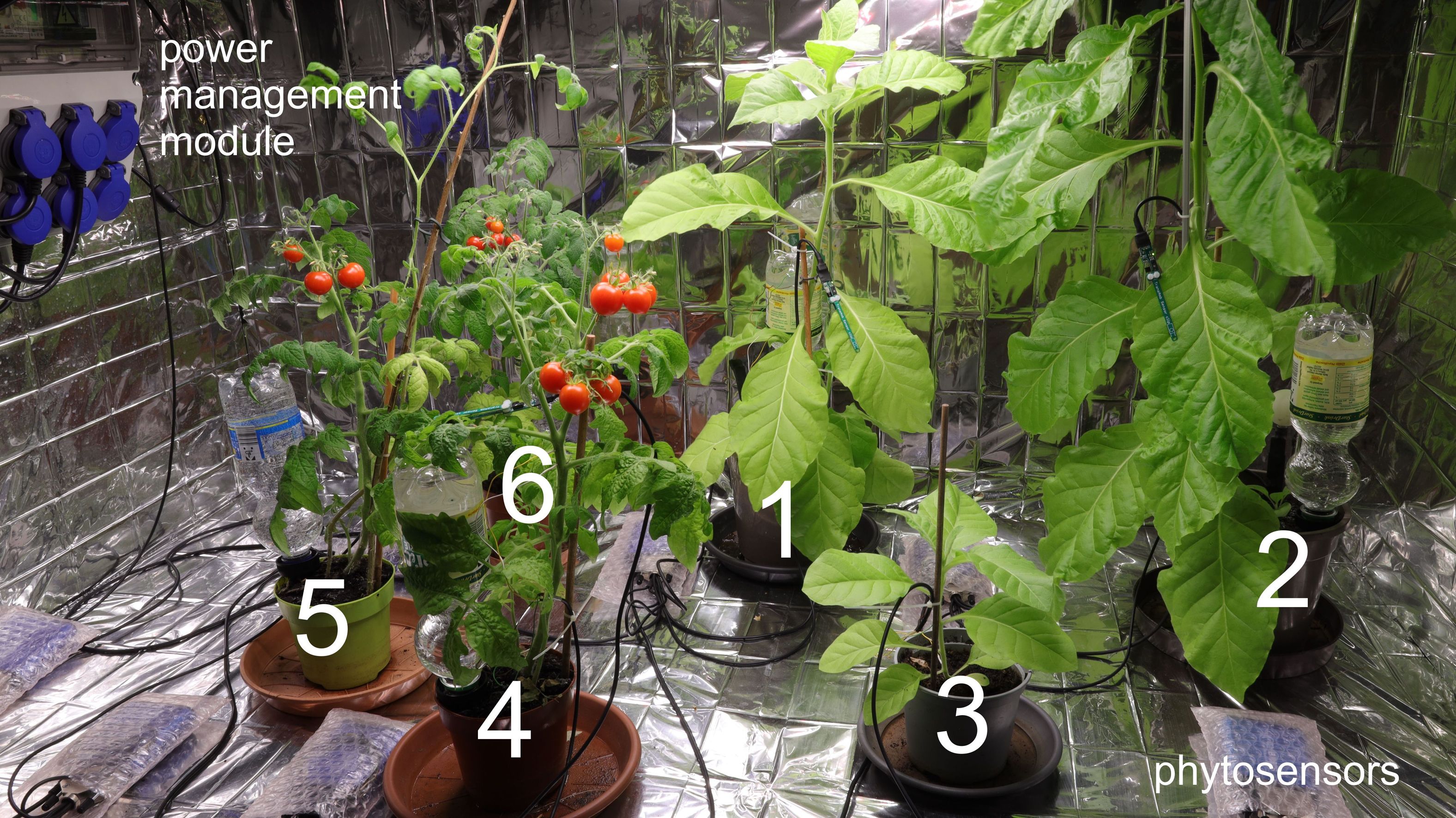}}
\caption{\small Measurement chamber with 3 tobacco and 3 tomato plants. Shown are phytosensors and the power management module for autonomous ON/OFF of light, aeration and preparing \ce{O_3}-air mix with the required \ce{O_3} concentration.\label{fig:experimentalSetup}}
\end{figure*}

\begin{figure}[htp]
\centering
\subfigure{\includegraphics[width=.49\textwidth]{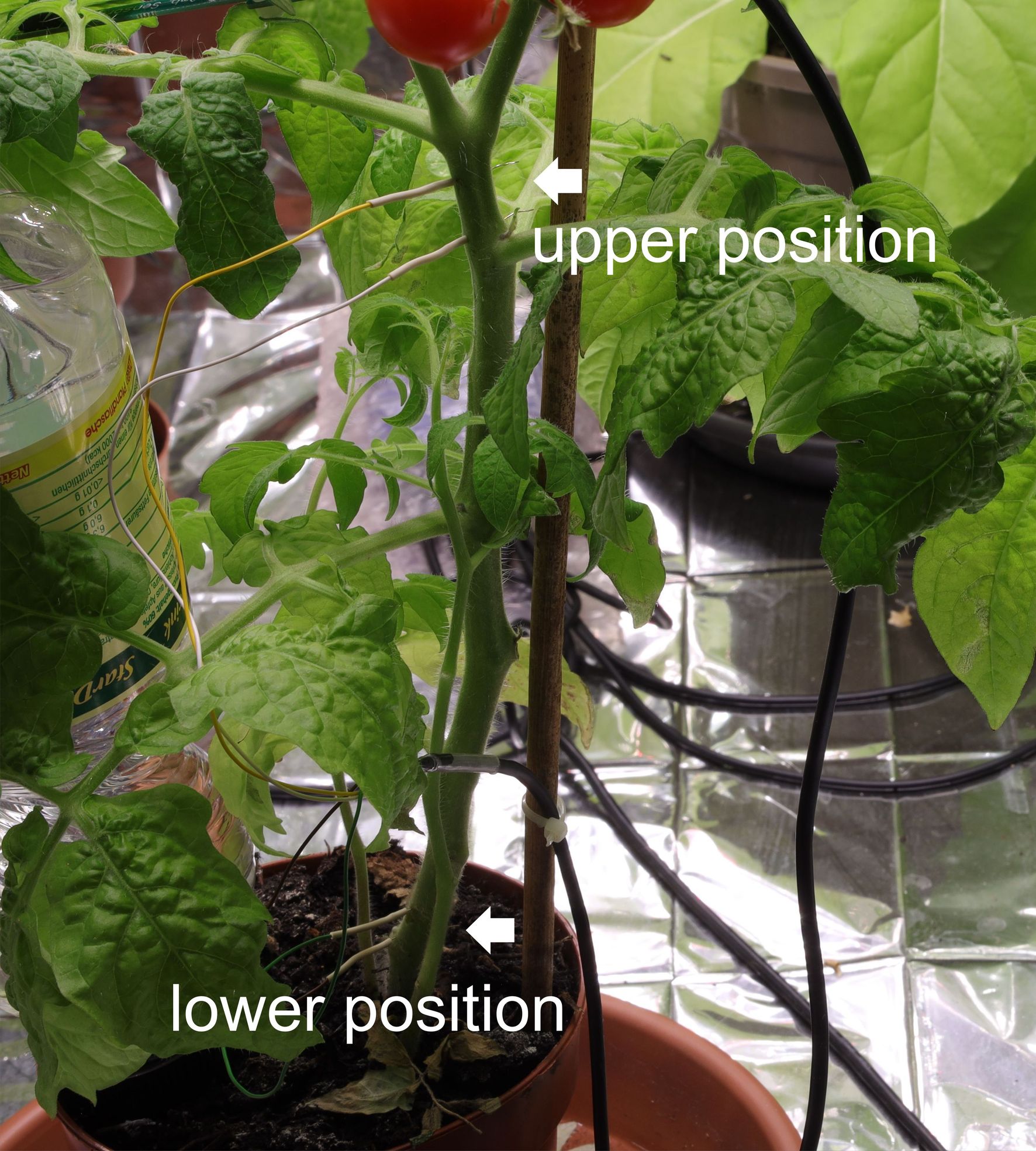}}
\caption{\small Placement of EIS electrodes with lower and upper positions. \label{fig:EIS_electrodes}}
\end{figure}

\begin{figure}[htp]
\centering
\subfigure{\includegraphics[width=.49\textwidth]{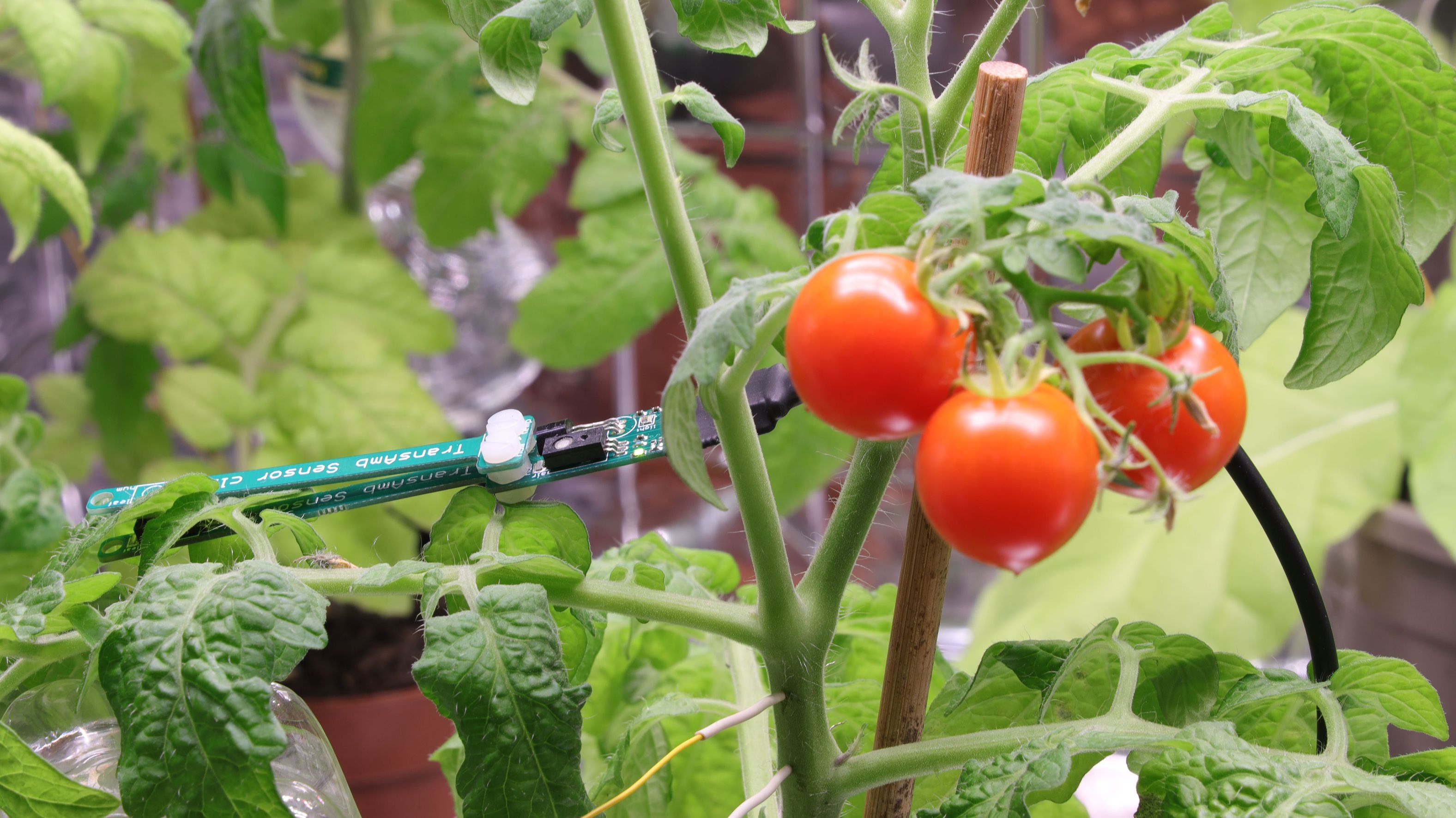}}
\caption{\small Placement of the transpiration sensor.\label{fig:Transpiration}}
\end{figure}

\textbf{Indoor setup.} Measurements have been conducted in a measurement chamber shown in Figs. \ref{fig:experimentalSetup}-\ref{fig:Transpiration} with three tobacco (\textit{Nicotiana tabacum, var. Habana}) and three tomato (\textit{Solanum lucopersicum} 'Balkonzauberer') plants. 
Preparation of \ce{O_3}-air mix with the required \ce{O_3} concentration is done in a 40L preparatory chamber with a low power \ce{O_3} generator (JB-OZ-S28 generator, 500mg/h in water part) and air pump (MT100, Air Cyrcle, diameter 100mm, 90 m$^3$/h), see the scheme in Fig. \ref{fig:O3_scheme}. To reduce the level of produced \ce{O_3}, the ozone generator is powered by 220V-PWM modulation with $<$0.1\% of duty cycles. 

The setup uses four different \ce{O_3} detectors/sensors: MQ131 Low (Winsen, 10--1000ppb, resolution: 1 ppb), ULPSM-O3 968-04 (SPEC, 0--20 ppm, resolution: 0.1ppm), DM509-O3 (0--5 ppm, resolution: 0.001 ppm), several measurements and calibrations have been done with POM (2B Technologies, 0-10 ppm, resolution: 1.5 ppb). Sensors are operated with 24 bit ADC in the data logger, and calibrated to a current level of atmospheric \ce{O_3}. Sec. \ref{sec:lowO3} describes the used approach for generation and sensing of low \ce{O_3} concentrations in more detail.

Preparatory measurements demonstrated a dependency on the irrigation (soil, hydroponics, substrate + hydroponics), finally the best solution for such measurements represents a microcapillary irrigation with soil-based growth. A full spectra LED light is used as a light source. To prevent the plants from overheating, the measuring chamber is open on one side and has two continuously operating 200 mm fans. Since the temperature and humidity should be the same before and during ozone exposure, no intervention is made in the setup during daylight hours.

\begin{figure}[htp]
\centering
\subfigure{\includegraphics[width=.49\textwidth]{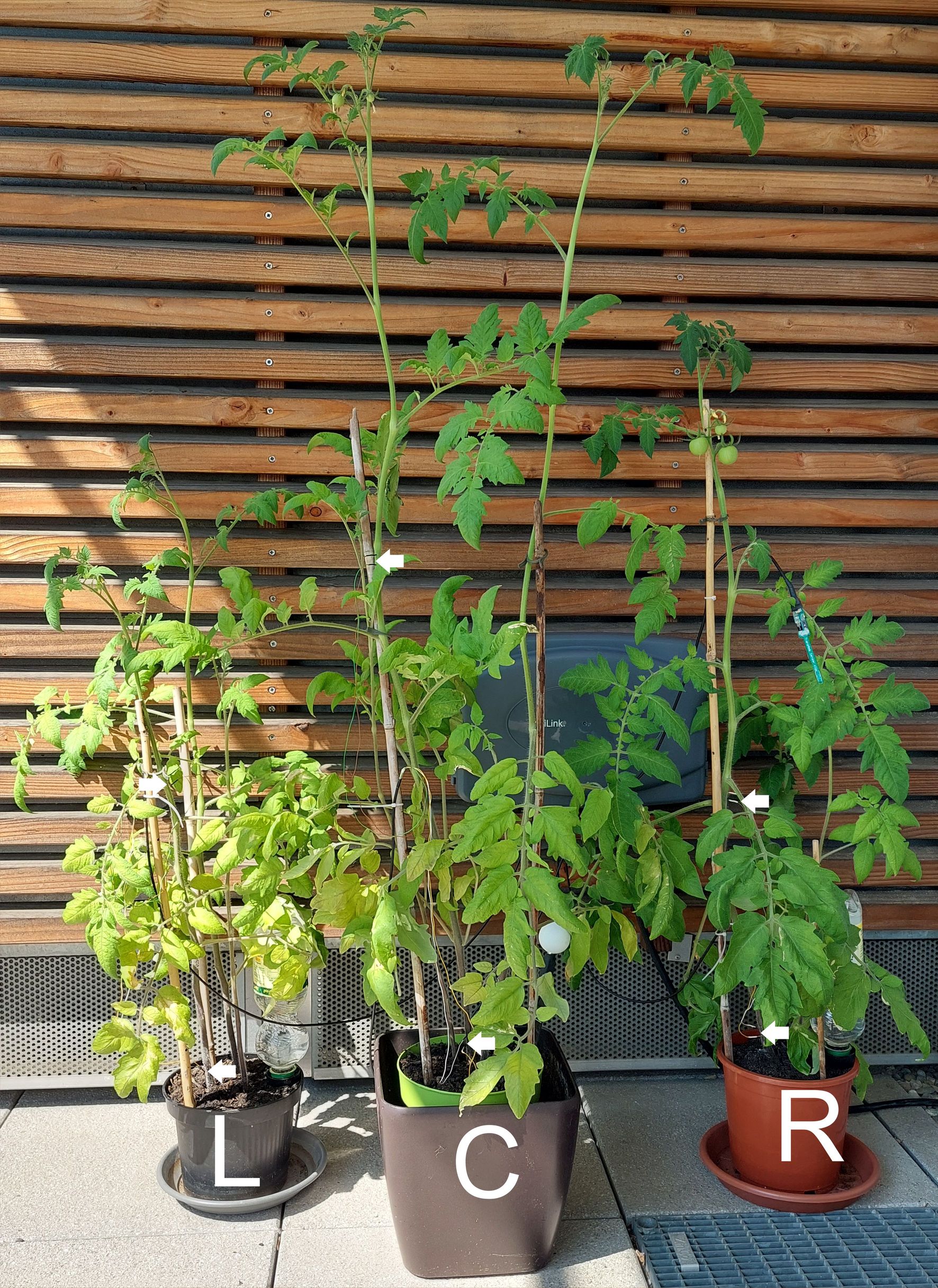}}
\caption{\small Outdoor setup with three tomato plants, arrows show lower and upper positions of EIS electrodes. Plants are placed in shadow however afternoon they are exposed to a direct sunlight, especially the L-plant. \label{fig:outdoorSetup}}
\end{figure}

\textbf{Outdoor setup.} Outdoor setup has three tomato plants (\textit{Solanum lucopersicum} 'Hellfruch') placed particularly in shadow to minimize heat stress, see Fig. \ref{fig:outdoorSetup}. L- and R-plants have microcapillary irrigation (however they needed additional irrigation due to hot summer season), C-plant has the hydroponic+substrate solution for irrigation. This setup has the same placement of sensors as the indoor setup.

Phytosensing in both setups is conducted by 6 (indoor) and 3 (outdoor) independent CYBRES phytosensing devices connected to PC via USB-to-USB isolators (to decrease noise and to avoid USB ground loops). In indoor setup, the power management module \cite{CYBAPP28} is connected to one of phytosensors and enables remote/autonomous switching of light, aeration and preparation of \ce{O_3}-air mix. 

Two physiological parameters are measured in all plants: tissues impedance in time-continuous mode with the fixed excitation frequency (450Hz) in two locations with lower and upper positions of electrodes, see Fig. \ref{fig:EIS_electrodes}, and leaf transpiration, see Fig. \ref{fig:Transpiration}. Additionally, soil moisture and temperature, air temperature and humidity, illumination, magnetic fields (by 3D magnetometer), air pressure and radio-frequency emission (450Mhz-2.5Ghz range) are recorded. Sampling rate is about 2 sps.

\subsection{Generation and detection of low concentration of \ce{O_3}}
\label{sec:lowO3}

High concentration of ozone ($>$1ppm) can be generated and detected by a number of different \ce{O_3} generators and sensors, it has an essential impact on plant physiology \cite{Buss23}. However, working with a low concentration of ozone (15-190 ppb over atmospheric \ce{O_3}) represents some challenges for both generating and sensing parts. Due to PWM-modulation, \ce{O_3} from the generator should be first diluted in a large amount of air (in a preparatory chamber) and then pumped to the measurement chamber with plants. Low concentration of \ce{O_3} has inhomogeneous distribution of \ce{O_3}-air mix, it needs specific strategies for control a proper concentration of ozone. Different polymers are susceptible to oxidation by ozone \cite{KEFELI1971904}, this creates additional local inequalities of \ce{O_3} distribution. Discovered issues on the sensing part are related to calibration of different \ce{O_3} sensors for low-scale measurements, inertially of sensor reaction on \ce{O_3} exposure and dependency to a measurement position due to inhomogeneous distribution of \ce{O_3}.

\begin{figure}[htp]
\centering
\subfigure{\includegraphics[width=.49\textwidth]{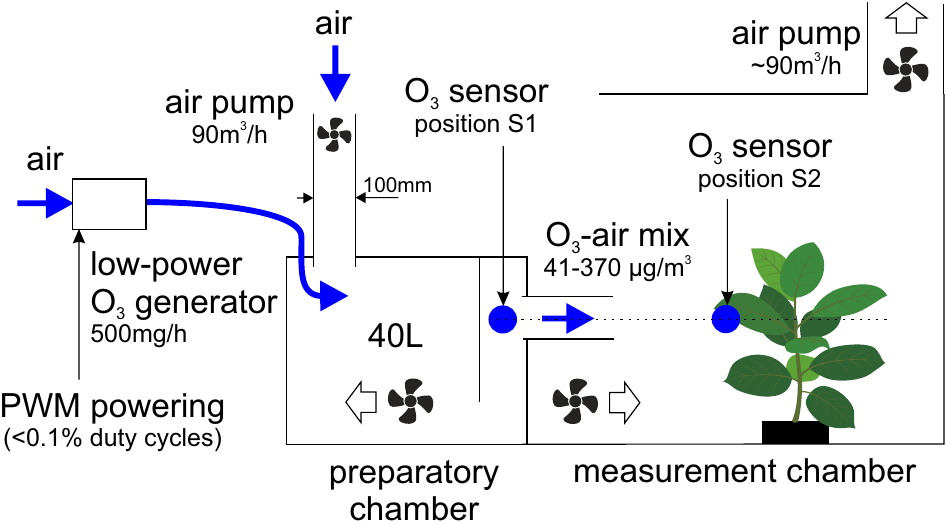}}
\caption{\small Scheme of experimental setup for generation and detection of low \ce{O_3} concentrations.
\label{fig:O3_scheme}}
\end{figure}	

The used scheme of \ce{O_3} generation and distribution is shown in Fig. \ref{fig:O3_scheme}. Outputs of the air pump and \ce{O_3} generator are fed to a 40L preparatory chamber. \ce{O_3} generator is powered by 220V-PWM modulation with $<$0.1\% of duty cycles (periodical on-off, e.g. on-time 2 sec., off-time 240 sec. -- 2/240). The \ce{O_3}-air mix from the preparatory chamber enters to the measurement chamber with plants. Air fans distribute the \ce{O_3}-air mix and remove it from the laboratory to avoid accumulation of \ce{O_3}. ON/OFF switching of all elements is conducted remotely or autonomously by the power management module, connected to one of phytosensor devices; temperature and humidity are constant during experimental attempts, see supplementary Fig. \ref{fig:exp_1508_add}.

\begin{figure}[htp]
\centering
\subfigure{\includegraphics[width=.49\textwidth]{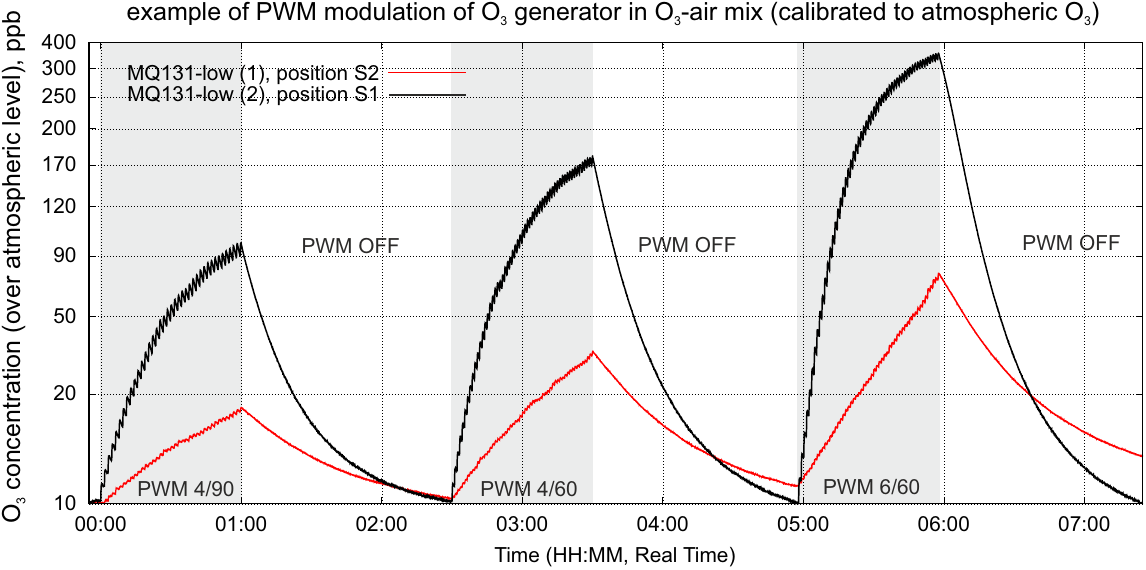}}
\caption{\small Examples of \ce{O_3} measurements and PWM modulation of \ce{O_3}-air mix in sensor positions S1 and S2 simultaneously by two equal \ce{O_3} sensors.
\label{fig:O3_test}}
\end{figure}		

Calibration of electrochemical \ce{O_3} sensors is performed based on their conversion diagrams to the level of atmospheric \ce{O_3}, see Fig. \ref{fig:O3_athmosphere}, considering humidity and temperature, and decomposition of \ce{O_3} to \ce{O_2} in laboratory \cite{BATAKLIEV14}. Additionally, the calibration from conversion diagrams was compared to a factory-calibration of DM509-O3 and POM meters. Based on these tests ULPSM-O3 sensor was removed from measurements and POM was used only for initial calibrations of the PWM-based control, see Table \ref{tab:PWM-O3}. Both MQ131 and DM509 connected to external 24 bit data logger provide a high resolution measurements in a low range of \ce{O_3} scale, however have different reaction time. \ce{O_3} generator produces ozone nonlinearly at short PWM impulses, see the coefficient $k_1$=1.4-0.77 in Table \ref{tab:PWM-O3}, with longer PWM durations it generates disproportionately more \ce{O_3}.  

\begin{table}[h]
\begin{center}
\caption{\small \ce{O_3} concentrations with different PWM settings of \ce{O_3} generator (500 mg/h) and constant air flow (90 m$^3$/h), conversion \ce{\mu g/m^3} to ppb at t=27C., due to PWM-modulation the measured values are rounded to integers. \label{tab:PWM-O3}}
\fontsize {9} {10} \selectfont
\begin{tabular}{
p{1.0cm}@{\extracolsep{3mm}}
p{1.0cm}@{\extracolsep{3mm}}
p{1.0cm}@{\extracolsep{3mm}}
p{1.0cm}@{\extracolsep{3mm}}
p{1.0cm}@{\extracolsep{3mm}}
p{1.0cm}@{\extracolsep{3mm}}
p{1.0cm}@{\extracolsep{3mm}}
}\hline \hline
PWM, sec/sec& expected \ce{O_3},\ce{\mu g/m^3} & expected ppb & measured S1, \ce{O_3}, ppb & $k_1$, exp./ meas. S1 & measured S2 \ce{O_3}, ppb & $k_2$, meas. S1/S2 \\\hline
4/90        &   246.9          & 126.7 & 90-110  & 1.4-1.15 & 15-20 & 6-5.5 \\ 
4/60        &   370.3          & 188.9 & 160-180 & 1.18-1.05 & 25-30 & 6.4-6  \\ 
6/60        &   555.5          & 284.4 & 320-370 & 0.89-0.77 & 50-70 & 6.4-5.3  \\       
12/60 			&   1111.1         & 570.1 & -- & -- & $>$200 & --\\ 
\hline         
\end{tabular}
\end{center}
\end{table}

Since the measurement chamber is not closed in long-term measurements, \ce{O_3} concentration in it is lower than the set value in the preparatory chamber. We use two strategies for preparation of \ce{O_3}-air mix -- by controlling \ce{O_3} in the preparatory or in measurement chambers (sensor positions S1 or S2). The first case guarantees that the e\ce{O_3} level does not overstep the set value, however ozone concentrations in the sensor position S2 are 6.4-5.3 times lower than in S1, see the coefficient $k_2$ in Table \ref{tab:PWM-O3} and Fig. \ref{fig:O3_test}. In the second case we use the pumping strategy with excessive \ce{O_3} in S1 position during the fixed time and measure the actual ozone concentration in the position S2. Both strategies have impact on the \ce{O_3} dynamics. In experiments we used the excessive strategy with measurements by two equal \ce{O_3} sensors on the plant level (the position S2).

Finally, the range of 10-200 ppb \ce{O_3} is divided into two main subranges, with max. 15-20 ppb and max. 50-70 ppb, in which most measurements were performed (\ce{O_3} exposure twice per day during 11:00-12:00 and 17:00-18:00). Note that the peak of \ce{O_3} exposure is only reached in a short period of time, see Figs. \ref{fig:O3_test}, \ref{fig:comparison_1}. The ambient level of \ce{O_3} and the third range $>$200 ppb are used as control cases I and II to estimate the probability of false-negative and false-positive responses of biosensors.

\subsection{Principles of EIS analysis in indoor setup}
\label{sec:EIsAnalysis}

\begin{figure}[htp]
\centering
\subfigure[\label{fig:T1_0408_1}]{\includegraphics[width=.49\textwidth]{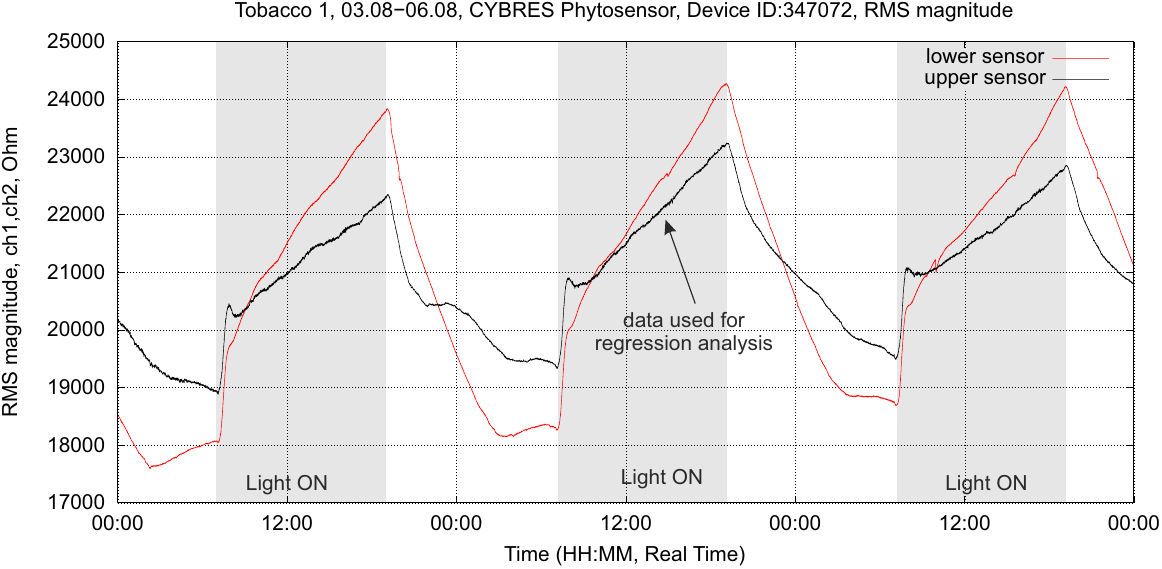}}
\subfigure[\label{fig:T1_0408_2}]{\includegraphics[width=.49\textwidth]{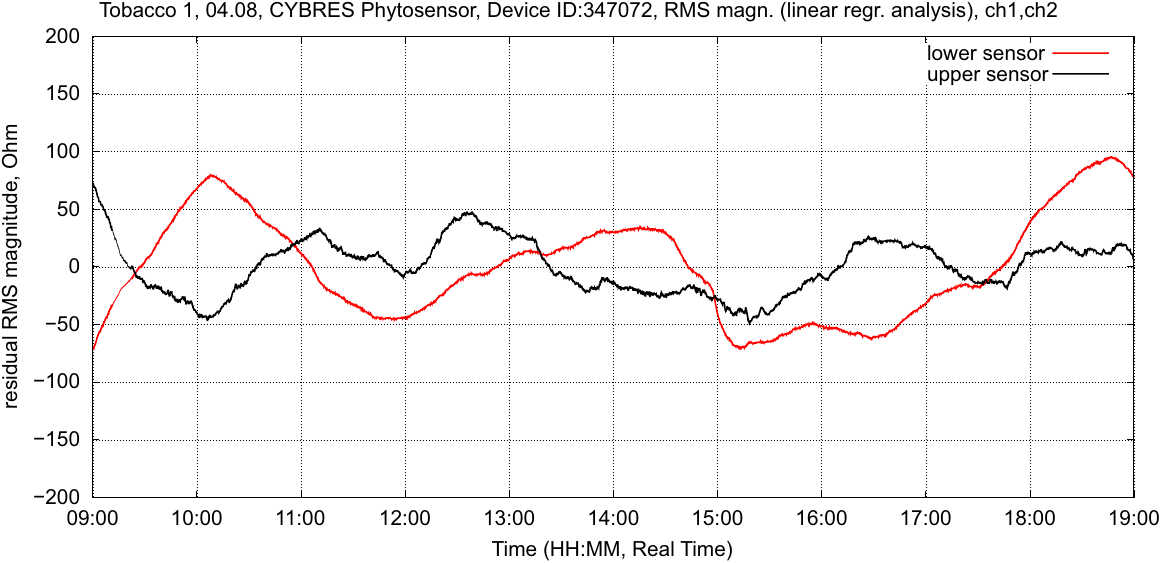}}
\caption{\small Example of EIS dynamics of upper and lower sensors: \textbf{(a)}  Original EIS dynamics during 72 hours of the tobacco plant 1; \textbf{(b)} The regression analysis of daily time during 10 hours. \label{fig:T1_0408}}
\end{figure}

Example of EIS data taken from the tobacco 1 (T1) plant during 72 hours is shown in Fig. \ref{fig:T1_0408_1}. Electrochemical dynamics is characterized by circadian rhythms indicating periodical movement of fluids in plant stem. An increase in electrochemical impedance means a decrease in the amount of ionic fluids in tissues (and vice versa). During the day we observe a slow decrease in fluid content through transpiration, while at night the roots supply the plant tissue with water (ionic aqueous solutions with dissolved minerals) and the fluid content increases. This is the simplest hydrodynamic model that explains the observed EIS dynamics, see \cite{kernbach2024Biohybrid} for more details. 

Day and night periods have a low noise dynamics (as long as the plants are not disturbed by additional stress), which can be processed by regression analysis. Fig. \ref{fig:T1_0408_2} shows an example of such an analysis for the interval 9.00--19.00 of a day period. We see that the 6000 Ohm variation is reduced to about 50 Ohm variation where all disturbances (e.g. applied \ce{O_3} stress) of the dynamics become detectable. 

Such an analysis imposes several requirements: 1) the increase of ozone in the \ce{O_3}-air mix should occur in a short time to achieve a significant change of electrochemical impedances; 2) the background period (without \ce{O_3} stress) should be long enough to provide good regression dynamics (periods without and with \ce{O_3} stress are typically in a 3:1 ratio); 3) plants should not be distorted by any other impacting factors such as temperature, humidity of irrigation, as they all affect the hydrodynamic system.  

\begin{figure}[htp]
\centering
\subfigure[\label{fig:T1_1208}]{\includegraphics[width=.49\textwidth]{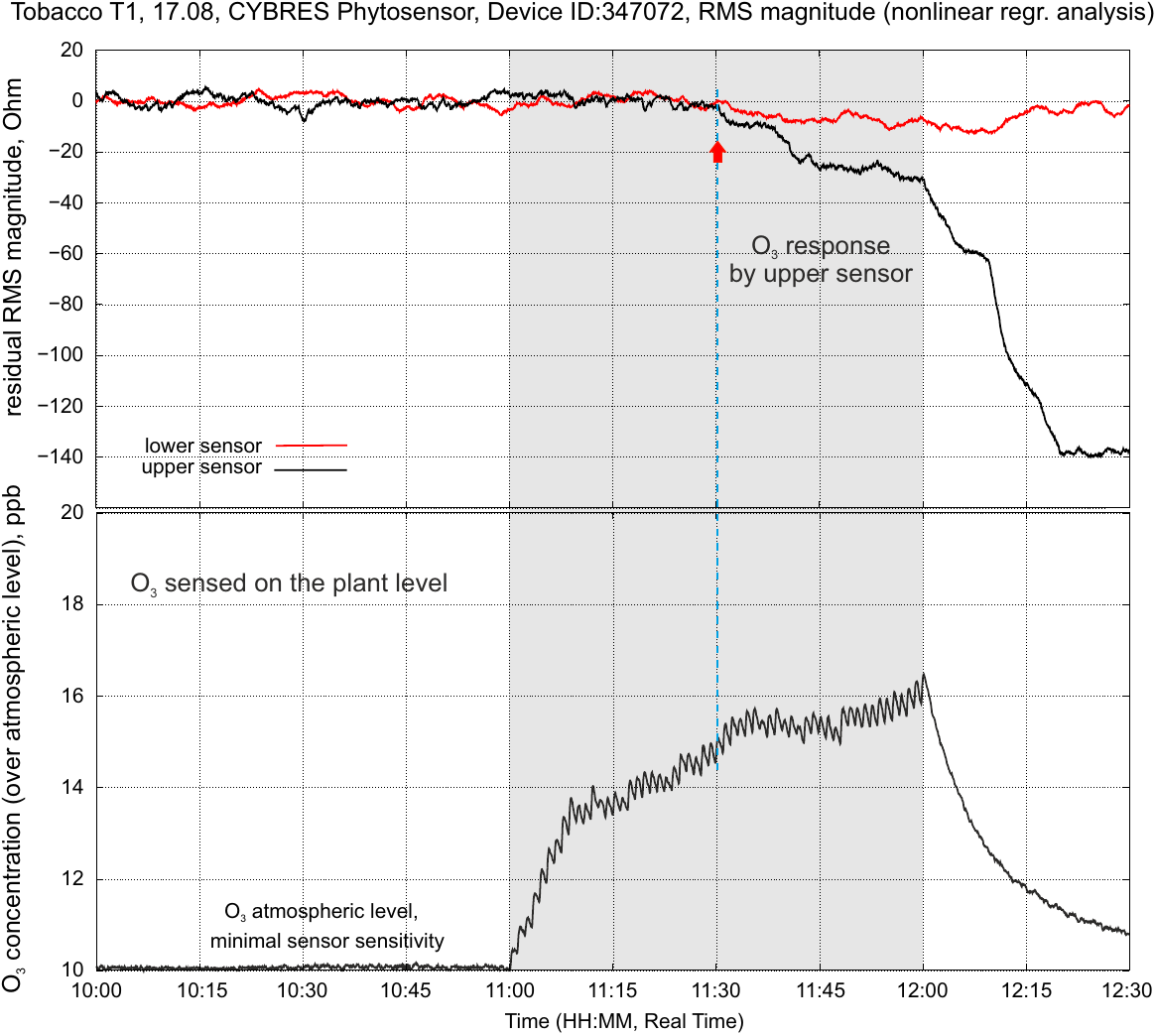}}
\subfigure[\label{fig:T4_inverseReaction}]{\includegraphics[width=.49\textwidth]{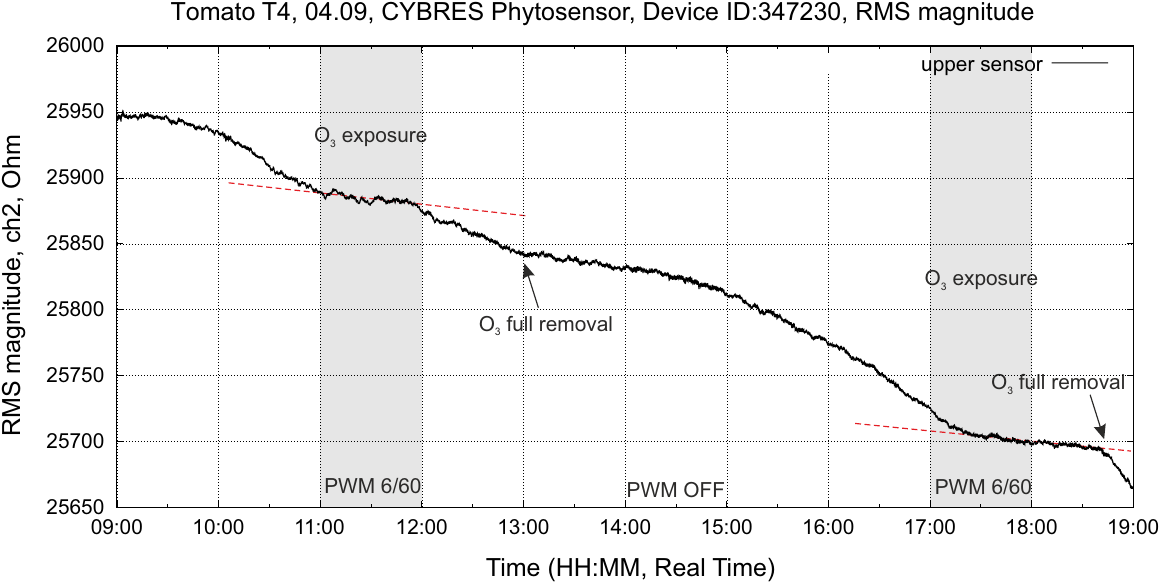}}
\caption{\small \textbf{(a)} Example of physiological response in the upper and lower positions of EIS sensors when exposed to low concentrations of \ce{O_3}; \textbf{(b)} The type II EIS dynamics of tomato plants.
}
\end{figure}

Fig. \ref{fig:T1_1208} shows the same plant with applied \ce{O_3} in the range of 15 ppb. Note that: 1) in most cases we only observe a change in electrochemical impedances in the upper sensor (increasing fluid content in the upper part of the stem associated with closing stomata); 2) the delay of biological response to applied \ce{O_3} is about 10–20 min.

\begin{figure}[htp]
\centering
\subfigure{\includegraphics[width=.49\textwidth]{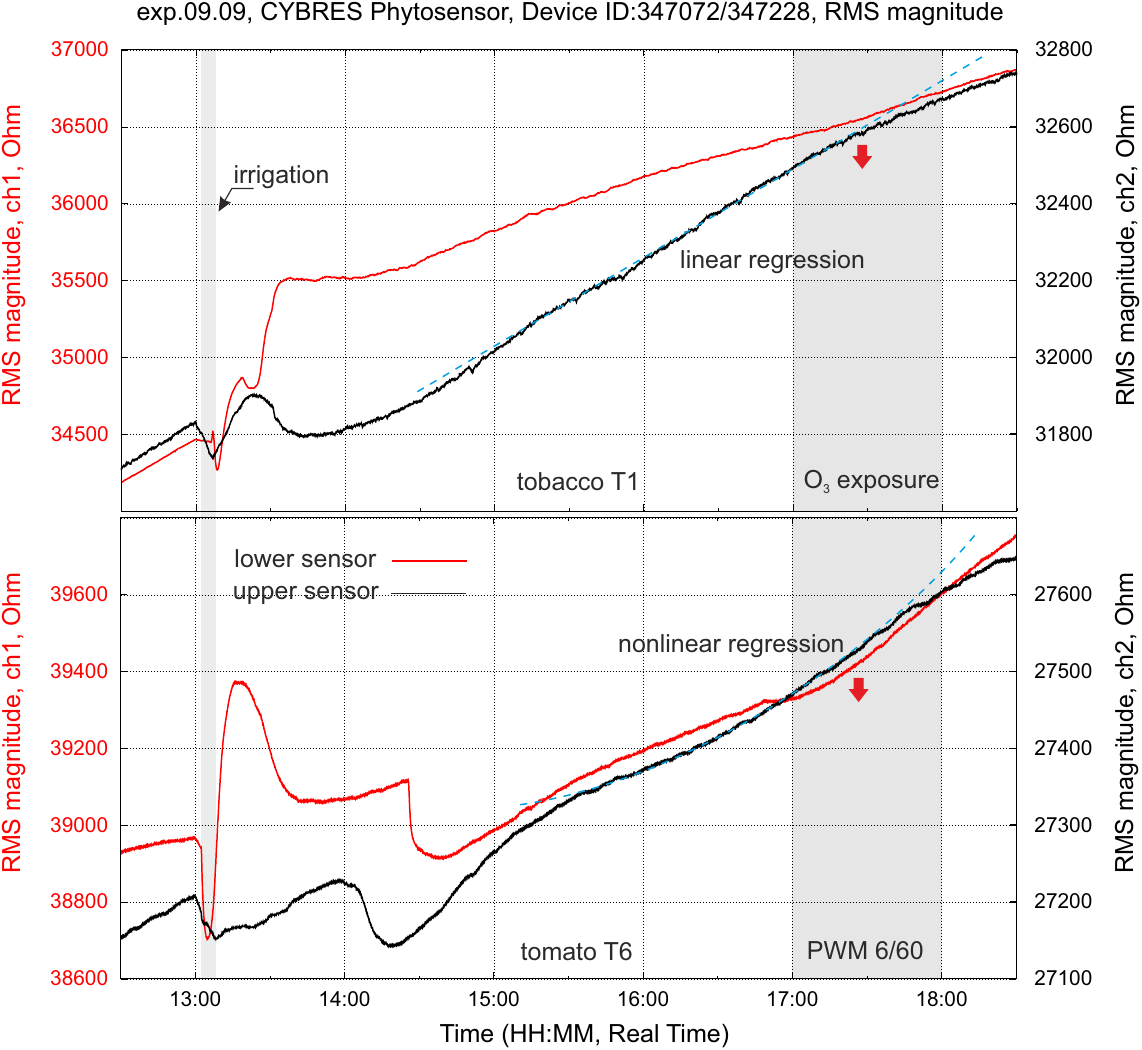}}
\caption{\small Interference between irrigation and \ce{O_3} exposure for tobacco and tomato plants. Both graphs exemplify also the linear and nonlinear regression used in the data analysis for the indoor setup. \label{irrigation_09_09}}
\end{figure}

Examples of linear and nonlinear regression are shown in Fig. \ref{irrigation_09_09}, their more detailed comparison with different regions of approximation and interpolation is shown in supplementary Fig. \ref{fig:exp_1508}, and explained in \cite{CYBRES_UserManual}. It is important to note that the change in EIS trends should be within the exposure range (i.e., there must be a clear correlation between \ce{O_3} exposure and a change in EIS trend). The regression analysis allows calculating different statistical or numerical parameters, see more in \cite{CYBAPP27,CYBAPP24}, which enable accurate characterization of imposed stress. 

Generally, we observe two types of EIS dynamics with increasing daily EIS trend as shown in Fig. \ref{fig:T1_0408_1}, further the type I, and decreasing daily EIS trend as shown in Fig. \ref{fig:T4_inverseReaction}, further the type II. The type I dynamics can be explained by decreasing fluid content due to transpiration, \ce{O_3} exposure reduces stomatal aperture and leads to decreasing EIS trend (higher fluid content in the stem) during exposure up to a full removal of \ce{O_3} from the lab. The regression analysis provides primarily a downward response for the type I. 

The type II dynamics is demonstrated mostly in stress conditions, for instance, the outdoor setup shows the type II dynamics in 90.6\% of all cases. During \ce{O_3} exposure, the EIS trend increases, the regression analysis demonstrates both upward and downward responses, see Fig. \ref{fig:comparison_2}. Considering different \ce{O_3}-protective reactions \cite{super2015cumulative}, \cite{ijms22126304}, here not only stomatal mechanisms are involved, but more complex hormonal regulation preventing toxic effects beyond the hydrodynamic system. Both types of EIS dynamics uses the same analysis, however their hypotheses should be tested during measurements.

The interference between irrigation and \ce{O_3} exposure is shown in Fig. \ref{irrigation_09_09} for tobacco and tomato plants. It initially distorts the hydrodynamic system near the roots and spreads along the stem to the upper parts. The disturbances last about 120–150 minutes. Taking into account the necessary background measurements, irrigation should be carried out at least 4–5 hours before \ce{O_3} exposure.      

\subsection{Principles of EIS analysis in outdoor setup}

There are two specific aspects of EIS dynamics that only occur outdoors: distortions produced by wind and rain, and artefacts produced by exposure to direct sunlight. If the distortion areas still reflect the hydrodynamics of fluid transportation, the regions with artefacts cannot be used for analysis. 

\begin{figure}[ht]
\centering
\subfigure{\includegraphics[width=.49\textwidth]{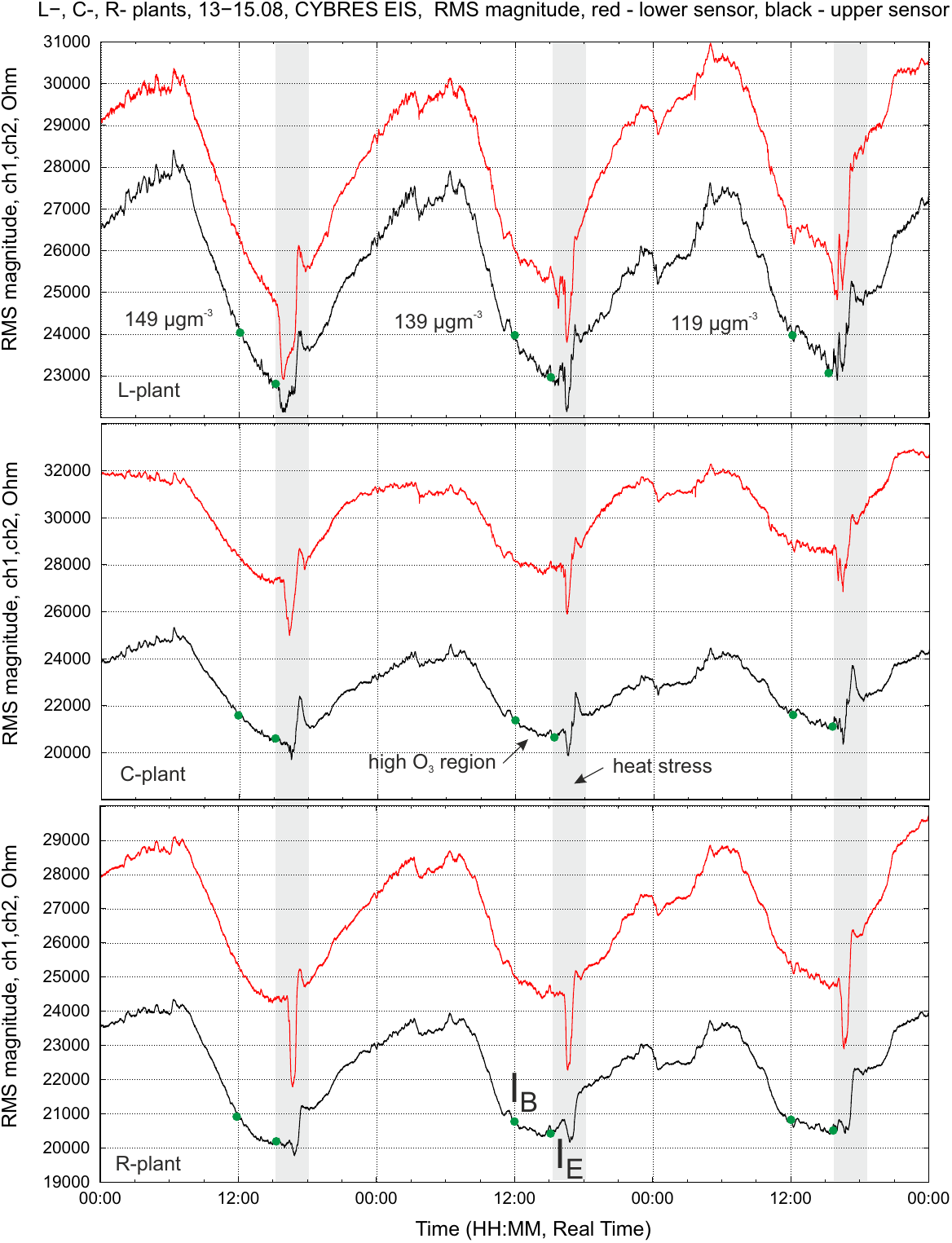}}
\caption{\small EIS dynamics from three tomato plants in outdoor setup during 72 hours, compare to Fig. \ref{fig:T1_0408} in indoor setup, shadow regions represent heat stress in afternoon time. Green points -- the values $I_B$ and $I_E$, see expression (\ref{eq:minMmax}), mark the region with the highest daily \ce{O_3} concentration. \label{fig:heatStress}}
\end{figure}

\begin{figure}[ht]
\centering
\subfigure[\label{fig:O3_athmosphere}]{\includegraphics[width=.49\textwidth]{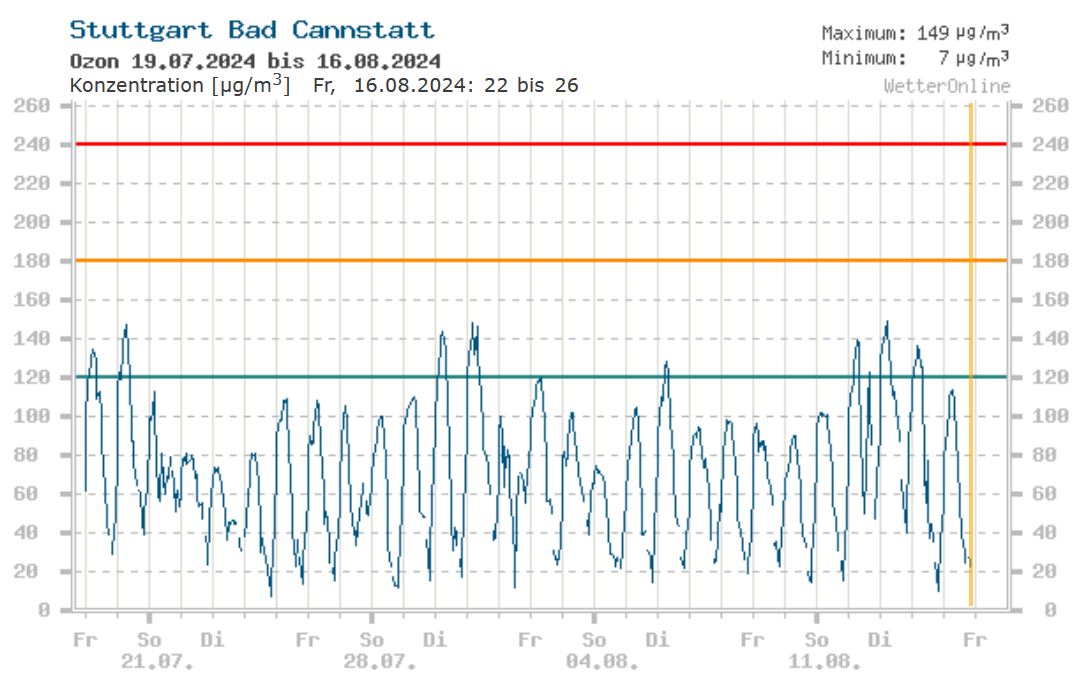}}
\subfigure[\label{fig:outdoorData_All}]{\includegraphics[width=.49\textwidth]{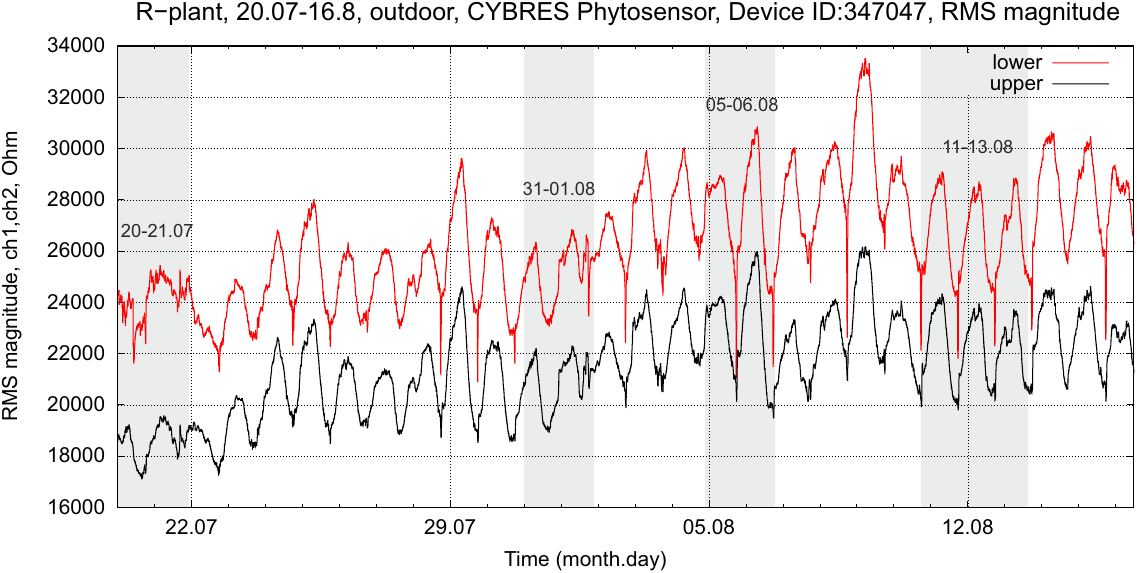}}
\caption{\small \textbf{(a)} Concentration of atmospheric \ce{O_3} during 27 days, data from the State Agency for the Environment Baden-W\"urttemberg (Landesanstalt f\"ur Umwelt Baden-W\"urttemberg), image from www.wetteronline.de/ozonwerte/stuttgart. \textbf{(b)} EIS dynamics the R-Plant, the high-ozone days are marked by grey areas}
\end{figure}

Fig. \ref{fig:heatStress} shows the EIS dynamics of three tomato plants outdoors over 72 hours. Shadow areas show the heat stress (as well as UV/IR radiation) in the afternoon. Corresponding peaks are largely present in both sensors of L-plant (this plant and both sensors are mostly in sunlight in the afternoon), R-plant has only short peaks in the lower sensor (which is partially in sunlight). Obviously, such peaks do not reflect the movement of fluids, but other phenomena related to heat and UV/IR radiation that affect plant physiology and physical chemistry of EIS sensing \cite{WANG2021100397}.

Another clearly visible effect is related to circadian rhythms -- the maximum impedance in indoor plants in Fig. \ref{fig:T1_0408_1} correlates with the end of lighting period (evening), while the maximum in outdoor plants is shifted to the morning (the type II of EIS dynamics). Taking into account that an increase in ozone leads to change of impedances in the upper sensor, we expect that EIS dynamics between 12.00 and 16.00 -- the points $I_B$ and $I_E$ in Fig. \ref{fig:heatStress} (beside artefact areas) on high-ozone days will have different contribution to 24-hours circadian rhythm compared to the same dynamics of low-ozone days.   

Since the used \ce{O_3} sensors are not intended for long-term outdoor measurements, we used data from two local air monitoring stations in outdoor measurements. Fig. \ref{fig:O3_athmosphere} shows a concentration of atmospheric \ce{O_3} during experiments from such a station, daily values range between 80 and 100 \ce{\mu gm^{-3}} that corresponds to 40.8 -- 50.9 ppb (1 ppb = 1.96 \ce{\mu gm^{-3}} at 25C), the largest value is 149 \ce{\mu gm^{-3}} or 75.9 ppb. Fig. \ref{fig:outdoorData_All} shows EIS dynamics of R-plant in the same period, the high-ozone days are marked by grey areas. The ideas of analysis can be based on compassion of $\Delta I=I_{B} - I_{E}$ values on high-\ce{O_3} and low-\ce{O_3} days. Ozone stress changes the inclination of $I_B$--$I_E$ region, we expect that numerical value of $\Delta I$ will reflect this impact. Since different circadian cycles are shifted to each other, it needs to analyse pairwise $\Delta I_{high O3}$ and $\Delta I_{low O3}$ from similar regions of long-term EIS dynamics. Next section demonstrates this analysis.

\section{Results}
\label{sec:Results}

\newcolumntype{g}{>{\columncolor{Gray}}}
\begin{table*}[h]
\begin{center}
\caption{\small Probability of a type I reaction upon exposure to \ce{O_3}, the indoor setup with T1-T6 plants, N -- the number of plant-sensor attempts, S$_L$ -- lower EIS sensor, S$_U$ -- upper EIS sensor. Cases 11:00-12:00 and 17:00-18:00 accumulate reactions of all plants in all days at these attempts. \label{tab:results}}
\fontsize {9} {10} \selectfont
\begin{tabular}{
p{0.3cm}@{\extracolsep{3mm}}
p{1.0cm}@{\extracolsep{3mm}}
p{1.0cm}@{\extracolsep{3mm}}
p{0.6cm}@{\extracolsep{3mm}}|
gp{0.5cm}@{\extracolsep{3mm}}
p{0.5cm}@{\extracolsep{3mm}}
gp{0.5cm}@{\extracolsep{3mm}}
p{0.5cm}@{\extracolsep{3mm}}
gp{0.5cm}@{\extracolsep{3mm}}
p{0.5cm}@{\extracolsep{3mm}}
gp{0.5cm}@{\extracolsep{3mm}}
p{0.5cm}@{\extracolsep{3mm}}
gp{0.5cm}@{\extracolsep{3mm}}
p{0.5cm}@{\extracolsep{3mm}}
gp{0.5cm}@{\extracolsep{3mm}}
p{0.6cm}@{\extracolsep{1mm}}|
p{0.5cm}@{\extracolsep{3mm}}
p{0.5cm}@{\extracolsep{3mm}}
p{0.5cm}@{\extracolsep{3mm}}
p{0.5cm}@{\extracolsep{3mm}}
}\hline \hline
  & PWM  &  \ce{O_3}, ppb  &   N   & S$_L$  & S$_U$  & S$_L$ & S$_U$ & S$_L$ & S$_U$ & S$_L$ & S$_U$ & S$_L$ & S$_U$ & S$_L$ & S$_U$ & S$_L$ & S$_U$ & S$_L$ & S$_U$\\\hline
	  &    &                 &       &\multicolumn{2}{c}{plant T1} & \multicolumn{2}{c}{plant T2} & \multicolumn{2}{c}{plant T3} & \multicolumn{2}{c}{plant T4} & \multicolumn{2}{c}{plant T5} & \multicolumn{2}{c}{plant T6} & \multicolumn{4}{l}{11:00-12:00~17:00-18:00} \\\hline
1 & control & ---	         &  180  & 0.20  & 0.13  & 0.20  & 0.13  & 0.27  & 0.13  & 0.07  & 0.07  & 0.13  & 0.20  & 0.13  & 0.13  & 0.19  & 0.14  & 0.15  & 0.13  \\   
2 & 4/90 &  15-20          &  204  & 0.71  & 0.65  & 0.47  & 0.76  & 0.35  & 0.53  & 0.53  & 0.47  & 0.53  & 0.76  & 0.53  & 0.88  & 0.40  & 0.73  & 0.63  & 0.63  \\ 	
3 & 6/60 &  50-70          &  252  & 0.43  & 0.76  & 0.33  & 0.86  & 0.52  & 0.62  & 0.33  & 0.57  & 0.38  & 0.86  & 0.52  & 0.90  & 0.32  & 0.80  & 0.52  & 0.73  \\ 
4 & 12/60 &  $>$200        &  132  & 0.27  & 0.64  & 0.73  & 0.91  & 0.91  & 0.91  & 0.55  & 0.73  & 0.18  & 0.91  & 0.36  & 0.82  & 0.53  & 0.93  & 0.43  & 0.80  \\	
\hline         
\end{tabular}
\end{center}
\end{table*}

\begin{figure}[ht]
\centering
\subfigure[\label{fig:comparison_1}]{\includegraphics[width=.49\textwidth]{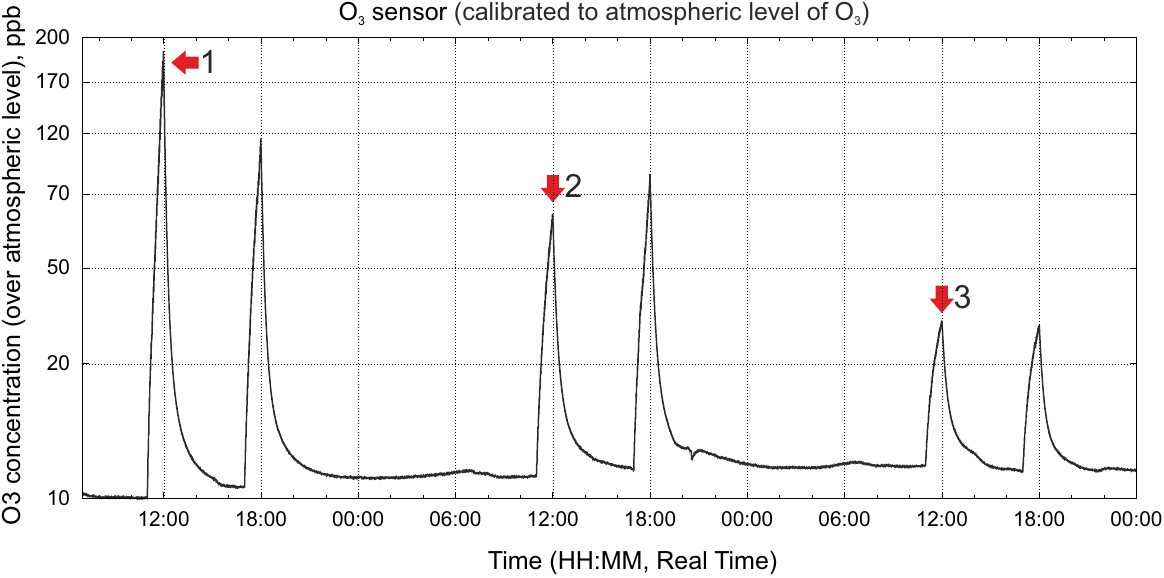}}
\subfigure[\label{fig:comparison_2}]{\includegraphics[width=.49\textwidth]{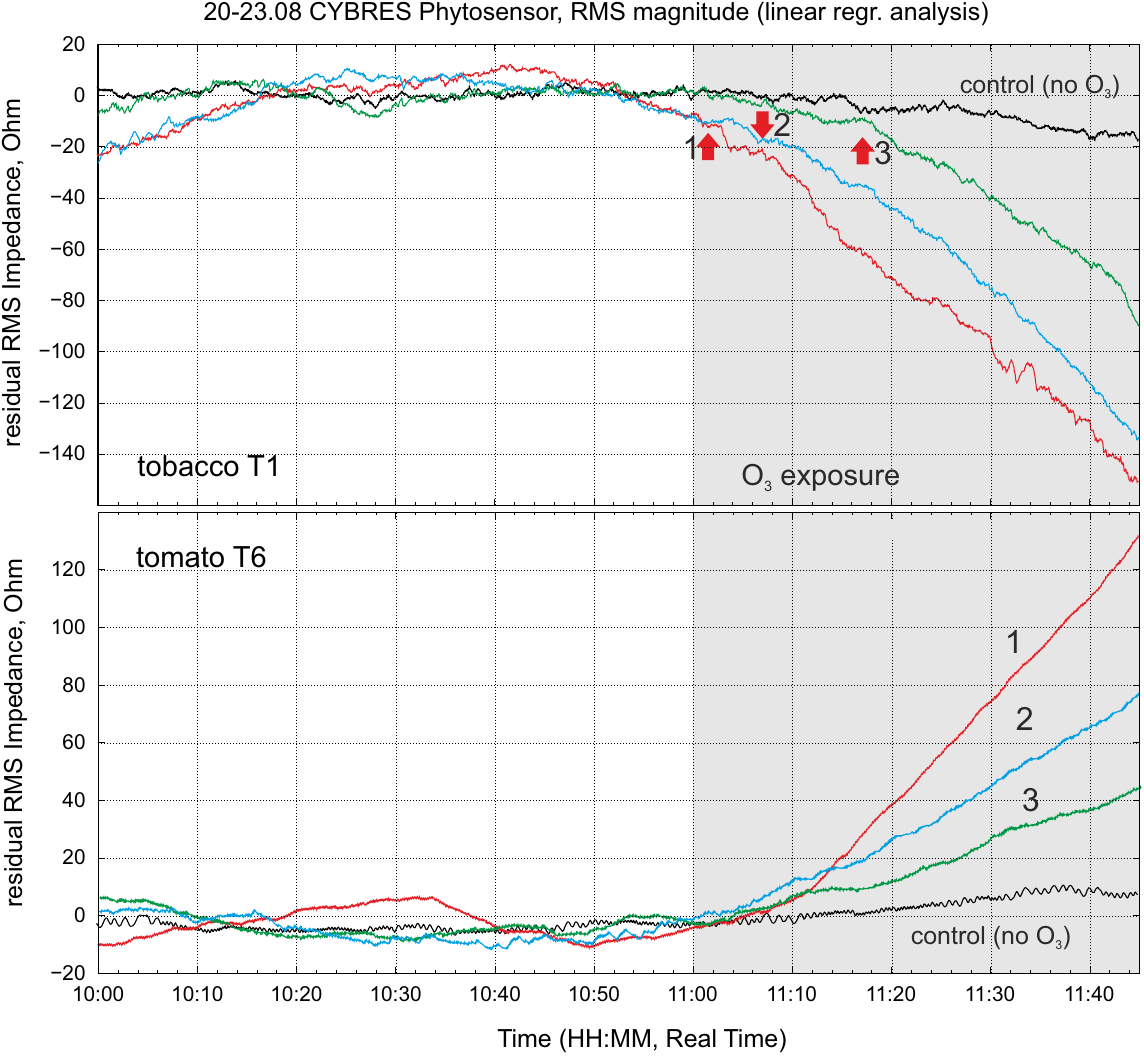}}
\caption{\small \textbf{(a)} Test with different concentrations of \ce{O_3}; \textbf{(b)} Reactions of tobacco and tomato plants (linear regression analysis): shift of inflection points and proportional inclination of EIS curves for both type I and type II responses.  
\label{fig:comparison}}
\end{figure}

\textbf{Indoor setup.} In general, the circadian EIS dynamics is specific for each plant species and environmental conditions, compare Figs. \ref{fig:heatStress} and \ref{fig:tomatoIndoor} for outdoor and indoor tomatoes and Fig. \ref{fig:T1_0408} for tobacco. However, all plants of the same species and in the same conditions demonstrated a homogeneous reaction on \ce{O_3} exposure, see supplementary Fig. \ref{fig:homogeneousTomato}. Comparing a reaction on different concentrations of \ce{O_3} (about 190 ppb, 60 ppb, 20 ppb), we see a small shift of inflection points (about 7-10 min in each case) and also proportional inclination of EIS curves for both type I and type II responses, see Fig. \ref{fig:comparison}. This demonstrates a possibility of qualitative but also quantitative measurements after calibration. We also note more evident reactions by tobacco with clear inflection points even for small concentrations of \ce{O_3} ($<$20ppb). 

Analysis is conducted automatically, for this the EIS dynamics is divided into 120 min of background region $B$ and 60 min of experimental region (\ce{O_3} exposure) $E$. Since the physiological reaction is delayed after the PWM is ON, $E$ is shifted by 30 min after the exposure starts. The original data $data(x)$ in the background region $B$ from EIS devices are approximated by the nonlinear function $fit_N(x)$ of 5th order, using the Levenberg-Marquardt algorithm \cite{Dennis83}, where we consider the residual curve
\begin{equation}
\label{eq:resifual}
R(x)=fit_{L,N}(x)-data(x).
\end{equation}
Finally, standard deviations $\sigma$ for R(x) curve are calculated; $\sigma_B$ characterizes the background, $\sigma_E$ characterizes the experimental regions. The ratio
\begin{equation}
\label{eq:psi}
\Psi=k\frac{\sigma_E}{\sigma_B}
\end{equation}
represents the final result: the more intense is the perturbations of the region $E$ in relation to $B$, the higher is the value of $\Psi$. The coefficient $k$ reflects the downward or upward trend, $k=-1$ if EIS dynamics goes down in the $E$ phase (less than zero) and $k=1$ -- otherwise. Since the EIS sensor has lower and upper channels, we calculate $\Psi_L$ and $\Psi_U$ for each experimental case. If $\Psi_U>0$ is received, the analysis is repeated by setting the $E$ region to the start of \ce{O_3} exposure, the lower from these results is collected in the Table \ref{tab:results}. This step allows avoiding artefacts of regression analysis caused by wrong recognition of physiological response. Additionally, the values $R_{E}(x)$ (value of $R(x)$ at the end of experimental region $E$) are also recorded. 

First, we test the hypotheses about type I and type II dynamics by analysing the $\Psi<0$ conditions. All \ce{O_3} exposures are coded by
\begin{eqnarray}
\label{eq:cond1}
1:& \Psi<0,\\
0:& \Psi>0. 
\end{eqnarray}  
All 0 and 1 are collected per plant, sensor and experimental series, and normalized to the [0..1] range. They can be considered as the probability of a type I reaction upon exposure to \ce{O_3}. Table \ref{tab:results} shows these results. We see that the type I dynamics represents $>$50\% for the upper sensor, plants T2, T5 and T6 have $>$75\% of type I dynamics, thus the condition (\ref{eq:cond1}) will be counted as the main sensor response.

The control series accumulate measurements conducted before experimental series and between them. Since approximation-interpolation approach will always deliver a small deviation, for the control cases we additionally require 
\begin{equation}
\label{eq:cond2}
|R_R(x)|>10,
\end{equation}
to avoid a wrong detection of low amplitude changes. In control measurements we observe the probability 0.07-0.27 (0.15$\pm0.06$) of false-positive response. The false-negative detections for the type I dynamics include both the type II hypothesis and non-responsive measurements -- we observe an increase in type II reactions in the last series with $>$200 ppb. The probability of false-negative detection can be roughly estimated from the $>$200 ppb case (upper sensor forenoon and afternoon) -- between 0.07 and 0.20. However, we emphasize that this is a specific form of false-negative results and can be improved by detecting biological reactions that include both type I and type II dynamics.

Both false-positive and false-negative reactions are explained by internal physiological dynamics, reactions to other stress factors, as well as by the regression analysis that requires a flat EIS dynamics in the $B$ region. Young plants, such as the T3 plant, have a lower level of reaction, their usage in phytosensing should be avoided. It needs also to pay attention on the high-performance biosensors, such as the tomato T6 plant, which has 0.88, 0.9 and 0.82 reactions on \ce{O_3} exposures.

\begin{figure}[ht]
\centering
\subfigure{\includegraphics[width=.49\textwidth]{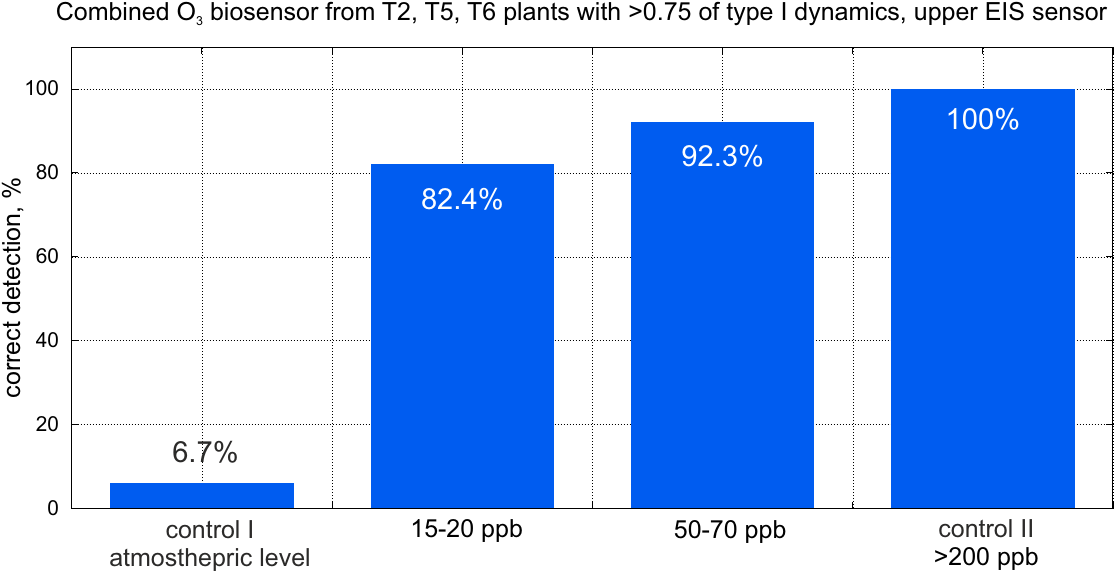}}
\caption{\small Combined sensor based on the type I dynamics -- combination of data (majority decision) from three phytosensors operating in parallel (T2, T5, T6 with $>$75\% of type I dynamics). \label{fig:combinedSensor}}
\end{figure}

We additionally analysed a combination of data from several plants, which improves a reliable identification of excessive \ce{O_3}. Such a combined biosensor represents the main practical interest. Three phytosensors operating in parallel (T2, T5, T6 with $>$75\% of type I dynamics), provide 0.82 and 0.92 reaction based on majority decision, see Fig. \ref{fig:combinedSensor}.

\begin{figure}[ht]
\centering
\subfigure{\includegraphics[width=.49\textwidth]{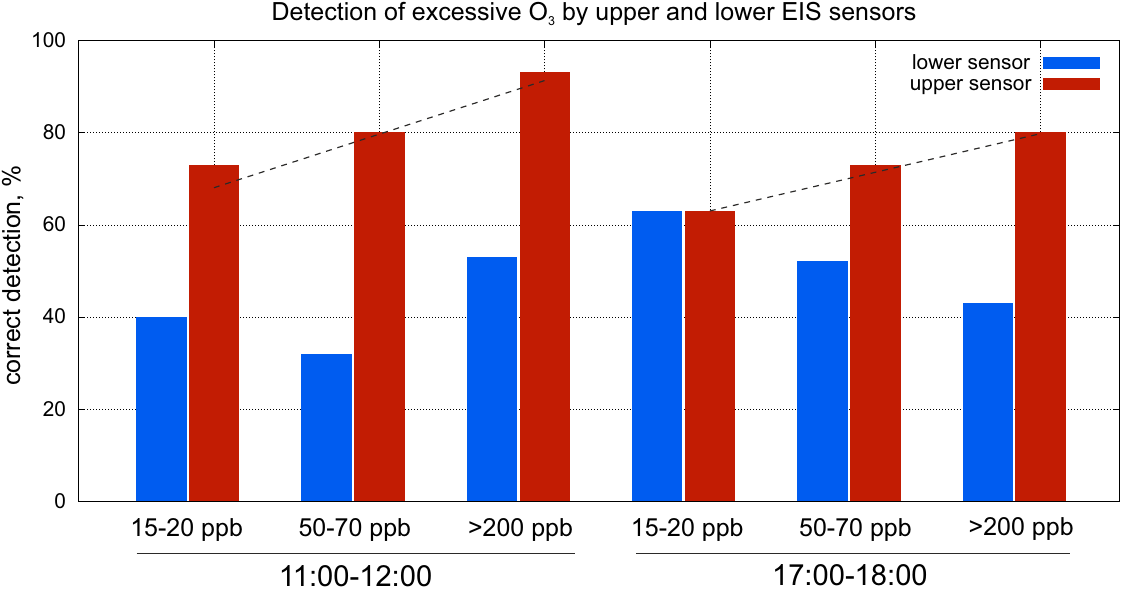}}
\caption{\small Detection of excessive O3 by upper and lower EIS sensors. \label{fig:combinedSensor2}}
\end{figure}

Another interesting result is related to detection of excessive \ce{O_3} by upper and lower EIS sensors, see Fig. \ref{fig:combinedSensor2}. Upper sensors provide better results for both 11:00-12:00 and 17:00-18:00 attempts. They also reflect the quantitative aspect shown in Fig. \ref{fig:comparison} -- an increase in \ce{O_3} leads to a stronger physiological response. We see that the position of electrodes plays an important role in the biosensor and reflects different physiological processes in the root and leaf area.

\textbf{Outdoor setup.} During 51 days of measurements (19.07-07.09) the local air monitoring station counted 15 days when the \ce{O_3} concentration overstepped 120 \ce{\mu gm^{-3}} (about 130 \ce{\mu gm^{-3}} in average). Due to possible \ce{O_3} variations between the pollution station and measurement location (about 3 km distance), we used data from two such stations (Stuttgart Bad-Cannstatt and Stuttgart Arnulf-Klett-Platz). We also selected 15 days with low \ce{O_3} concentration (about 80 \ce{\mu gm^{-3}} in average). Minimal difference between low-ozone and high-ozone days is about 40-50 \ce{\mu gm^{-3}}, which is comparable to the case 2 in indoor setup with 15-20 ppb range. These two data sets represent control and experimental cases. 

\begin{figure}[ht]
\centering
\subfigure{\includegraphics[width=.49\textwidth]{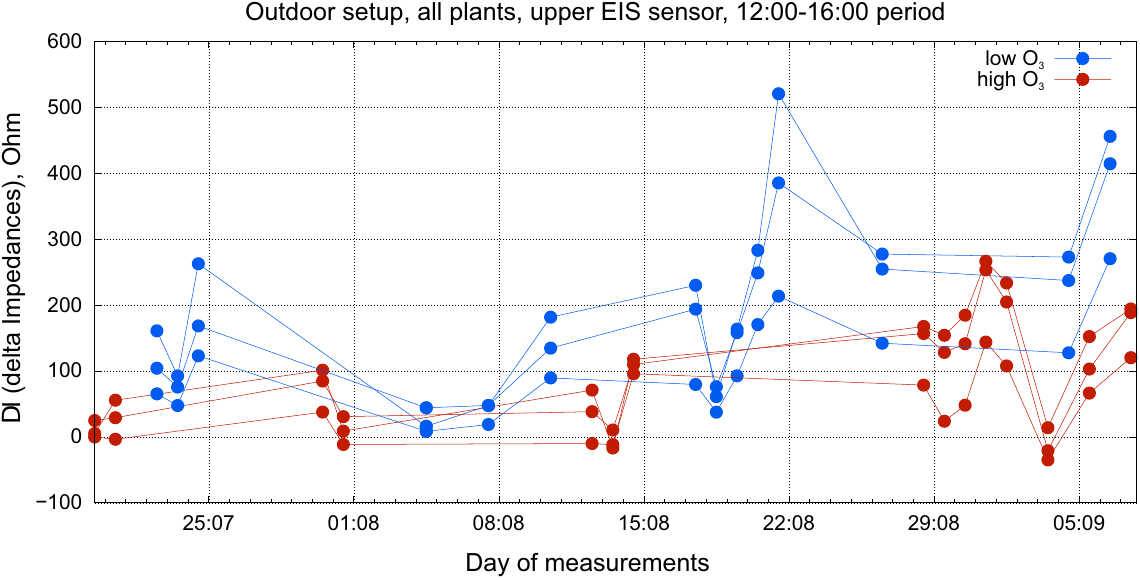}}
\caption{\small Values of $\Delta I$ from all three plant in outdoor setup, upper position of EIS sensors, 12:00-16:00 region of measurements, black and red points are low/high \ce{O_3} days, used in measurements \label{fig:outdoor_stats1}}
\end{figure}

For analysis we tested first the already discussed approach with the type II dynamics   
\begin{equation}
\label{eq:minMmax}
\Delta I=I_{B} - I_{E}>0,
\end{equation} 
where $I_{B}$, $I_{E}$ impedances of two regions of circadian rhythm at the begin and end of measurements (tested are three settings for $I_{B}$-$I_{E}$ as 12:00-16:00, 12:00-17:00 and 12:00-18:00). From 180 measurements (30 high/low \ce{O_3} days x 3 plants x 2 sensor locations), only 9.4\% are type I EIS dynamics, and 90.6\% -- type II dynamics. $I_{B}$, $I_{E}$ are calculated based on the mean of 10 EIS points. $\Delta I$ from all three plant in the outdoor setup (upper position of EIS sensors, 12:00-16:00 region of measurements) are shown in Fig. \ref{fig:outdoor_stats1}.

\begin{figure}[ht]
\centering
\subfigure{\includegraphics[width=.49\textwidth]{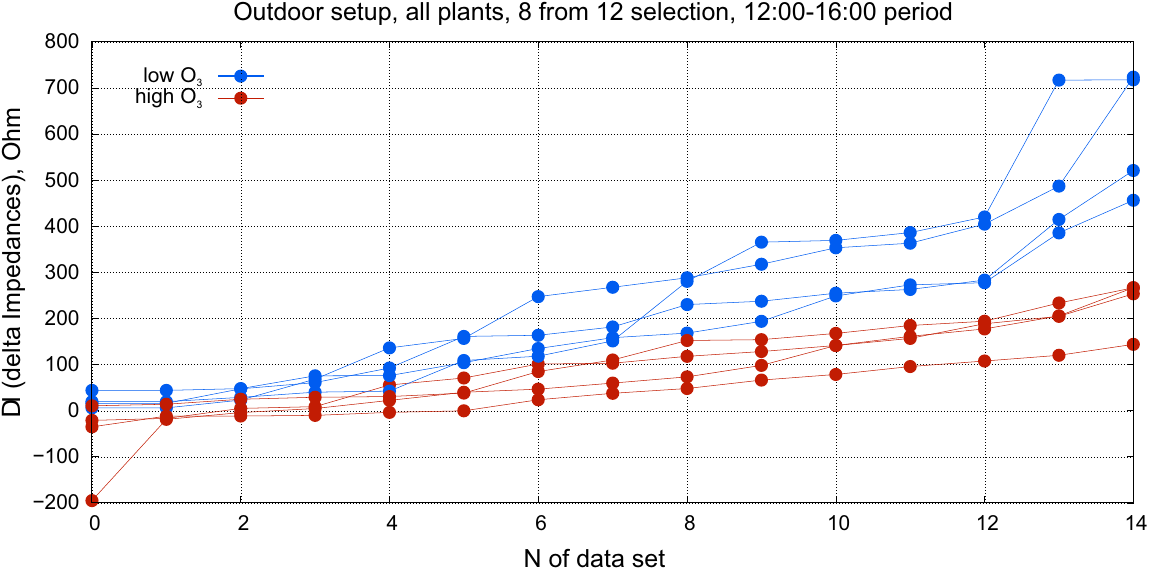}}
\caption{\small Separability of high/low \ce{O_3} sets for the selection 8 of 12 EIS data -- combination of measurements from different sensors and plants (measurements points are sorted by increasing values). \label{fig:outdoor_stats5}}
\end{figure}

Since the dynamics of $\Delta I$ is not normalized (it includes trend from different stressors), we apply the analysis of means for a combination of EIS sensors from different plants, Fig. \ref{fig:outdoor_stats5} demonstrates a separability of high/low \ce{O_3} sets for the selection 8 of 12 EIS data (3 plants x 2 EIS sensors x 2 high/low \ce{O_3} sets). Here we apply the same approach that is used in the indoor setup -- a combination of different sensors and different plants can improve the overall quality of sensing. Separability is one of the most important practical results because it enables the application of different pattern recognition approaches used in machine learning.

\begin{figure}[ht]
\centering
\subfigure{\includegraphics[width=.49\textwidth]{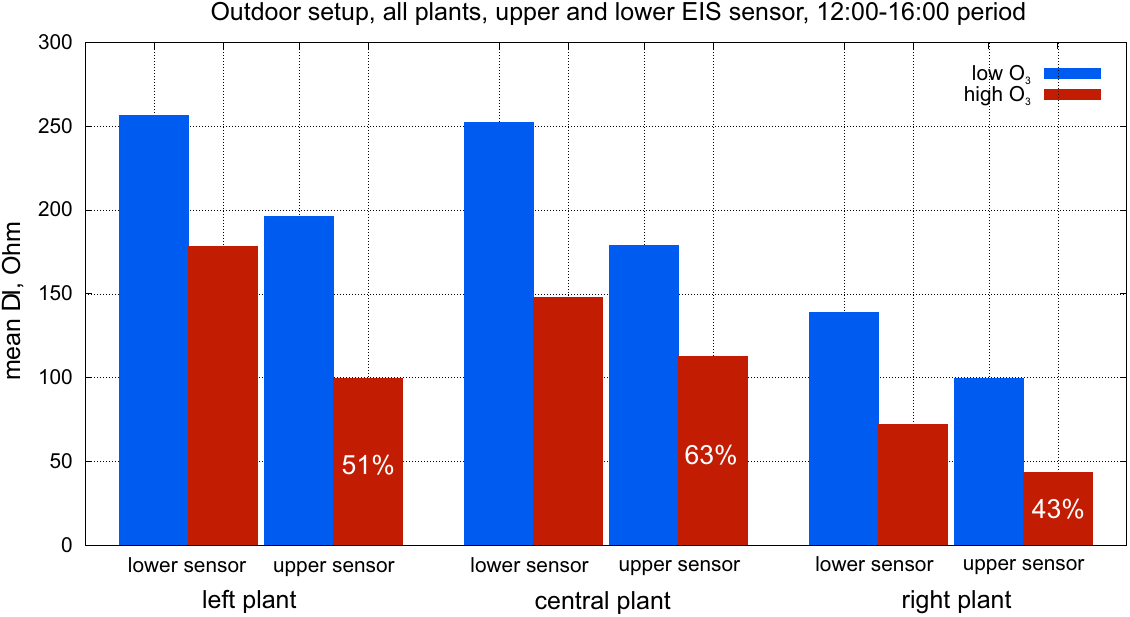}}
\caption{\small Comparison of means for all sensors and plants for 12:00-16:00 measurement region, numbers show the relation between high/low \ce{O_3} recognition by the same sensor, for instance, the low to high ration 0.43 (or high to low ratio 2.33) is obtained for the R-plant and upper EIS sensor. \label{fig:outdoor_stats2}}
\end{figure}

Finally, a comparison of means for all sensors and plants is shown in Fig. \ref{fig:outdoor_stats2}. We can see that the best ratio for high/low \ce{O_3} recognition -- high to low ratio 2.32 or low to high ratio 0.43 -- it obtained for the R-plant (the plant in shadow with minimal additional stress) and upper EIS sensor. Note that selecting $I_{B}$-$I_{E}$ as 12:00-17:00 or 12:00-18:00 that include heat and IR/UV impacts, see supplementary Fig. \ref{fig:outdoor_stats3}, the dynamics of $\Delta I$ receives large random components and the expression (\ref{eq:minMmax}) is no longer consistent. This demonstrates the importance of protecting measurement plants from direct sun radiation in outdoor conditions. Concluding the analysis of outdoor setup, we observe that a high \ce{O_3} concentration decreases the amplitude of EIS dynamics, which is also consistent with the type II dynamics of tomato plants in the indoor setup, see Fig. \ref{fig:T4_inverseReaction}.  

\section{Discussions}
\label{sec:Discussions}

One of the main methodological issues concerns type I vs. type II dynamics. In the selected analytical approach, we first test the hypothesis and then analyze the exposed EIS dynamics of \ce{O_3} based on this test. This leads to incorrect inclusion of 9\% – 25\% (outdoor--indoor) of correct results in false-negative cases. The switching between type I and type II dynamics has physiological reasons that need to be investigated in more detail. Applying other types of analysis, e.g. based on neural networks, can potentially improve the level of correct detections.

Another point of discussion in the indoor setup is related to the two-step dilution scheme of \ce{O_3} to 41-370 \ce{\mu g/m^3} from 500 \ce{mg/h} generator. The first step is isolated from the setup in the preparatory chamber, so that we can guarantee that high \ce{O_3} concentrations will not affect the plants even in a short time. However, the second step between sensor positions S1 and S2 is conducted in the measurement chamber with plants. Potentially, local \ce{O_3} inhomogeneity can reach some plants for a short time (in term of a few seconds considering two continuously operating 200m fans). Since EIS dynamics demonstrates a response in 10-20 minutes after the onset of exposure, we argue that such potential short term \ce{O_3} inhomogeneities will not essentially affect the EIS dynamics. 

The \ce{O_3} concentration has a typical pumping behaviour, from the atmospheric level up to the maximal value as shown in Fig. \ref{fig:O3_test}. Moreover, due to the \ce{O_3} to \ce{O_2} decomposition, the laboratory level of \ce{O_3} (e.g. reached over night) is lower than the environmental concentration. The lowest range calibration of all used sensors and meters has a specific absolute inaccuracy at beginning of the scale despise very precise relative measurements, moreover both \ce{O_3}-sensors and \ce{O_3}-generator have a nonlinearity at lower \ce{O_3} values. Considering these points, we can speak only about ranges of \ce{O_3} exposure over the current atmospheric level. This situation reflects the outdoor exposure, where \ce{O_3} level has a sinusoidal daily dynamics and typically we refer to a maximal \ce{O_3} level reached on that day.

One of most serious issues in the outdoor setup is related to overlapping of different stress factors (wind, rain, heat, UV/IR radiation) as shown in Fig. \ref{fig:heatStress}. Some of these factors affect the whole physiology, other only EIS sensing processes on the level of physical chemistry (in participial, the UV/IR radiation \cite{WANG2021100397}). Since the outdoor setup shown in Fig. \ref{fig:outdoorSetup} has a well-defined exposure time to direct sunlight, we remove these intervals for each plant individually. More generally, if plants are used as biosensors for environmental \ce{O_3} (or any other pollutant), they have to be protected from additional stress factors (or at least from a direct sunlight). This is also related to environmental sensors, for instance temperature/humidity sensors should not be exposed to a direct sunlight.

A separate point of discussion is the selectivity of biosensors to a specific environmental pollutant. In conducted experiments, we also noted a physiological response to PM (particulate matter), moreover, \ce{O_3} is correlated with nitrogen oxides (NOx) \cite{Souza17}. If we consider nonspecific physiological responses of plants at the hydrodynamic level, we can refer EIS-based biosensors to sensors that detect a biological response to complex environmental pollution from various sources. 

\begin{figure}[htb]
\centering
\subfigure[\label{fig:microgreens_1}]{\includegraphics[width=.49\textwidth]{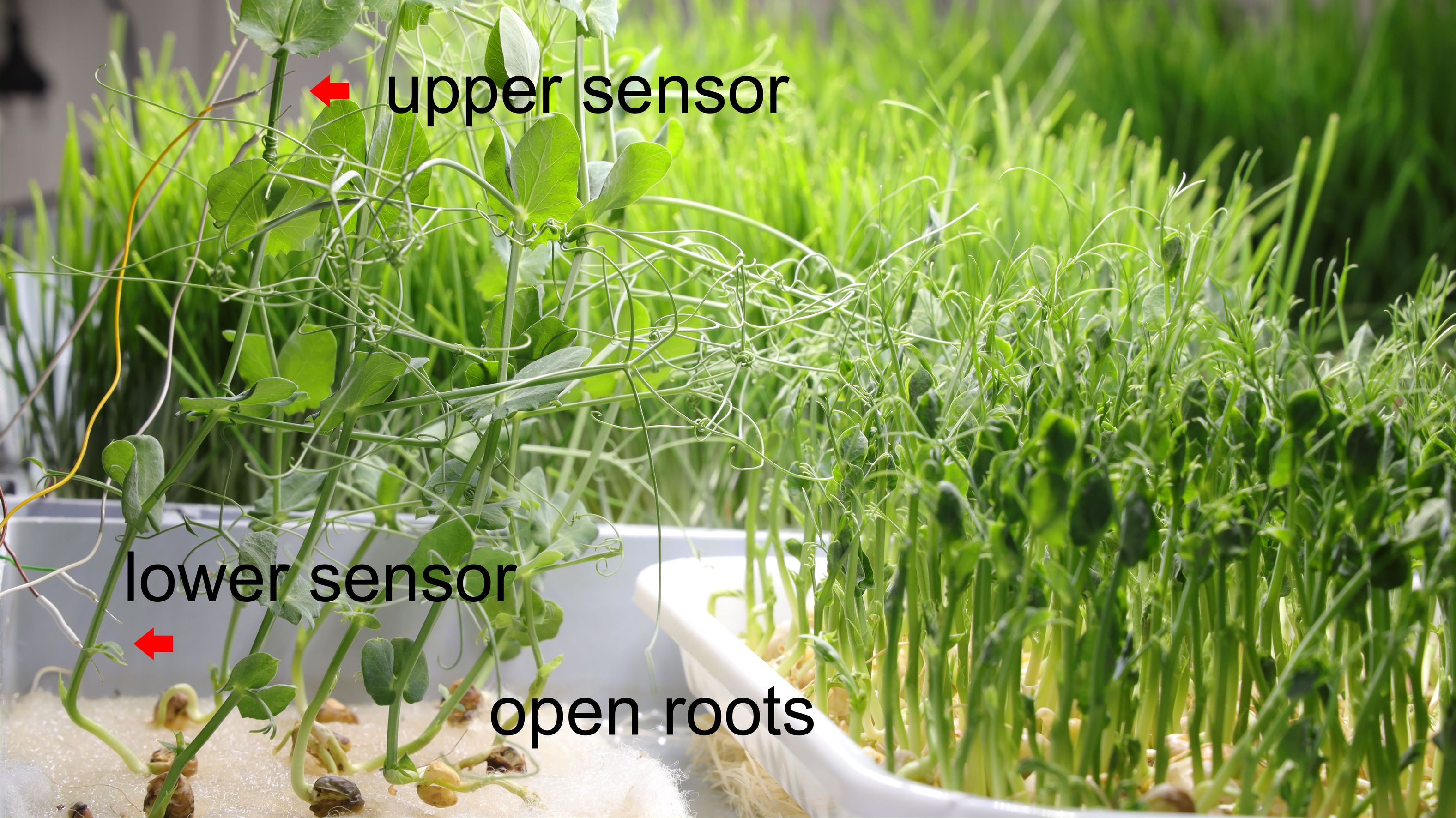}}
\subfigure[\label{fig:microgreens_2}]{\includegraphics[width=.49\textwidth]{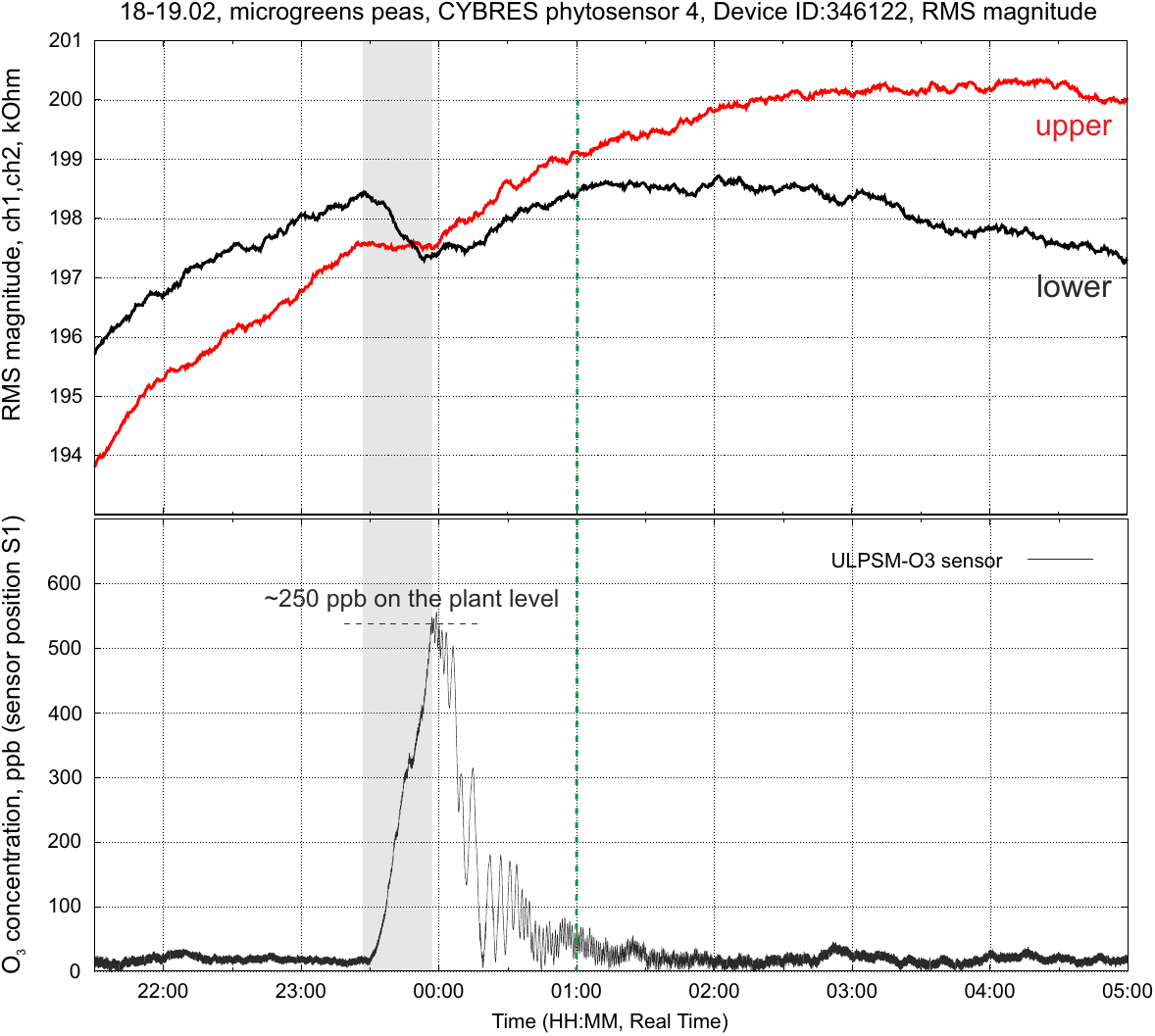}}
\caption{\small \textbf{(a)} Example of pea production in vertical farm without soil/substrate with open roots; \textbf{(b)} the response of EIS dynamics on \ce{O_3} exposure.
\label{fig:microgreens}}
\end{figure}

Interesting observation is made within preliminary tests with microgreens in vertical farm, see Fig. \ref{fig:microgreens_1}. As is typical for wheat and peas, their cultivation is conducted without soil/substrate with open roots and frequent periodical irrigation \cite{kernbach2024Biohybrid}. Exposure to \ce{O_3} demonstrated a higher response of the lower sensor, indicating a reaction of roots on ozone, see Fig. \ref{fig:microgreens_2}. Given the limited information on the mechanism of action of ozone on roots and the ongoing discussion \cite{Chen16}, such setups can be used to systematically investigate the impacts of environmental pollutants on new agricultural systems.

\section{Conclusion}

In this work we demonstrated a biological detection of a low concentration of \ce{O_3} by measuring electrochemical impedances of tissues in tobacco and tomato plants. Two setups have been created focused on isolating only \ce{O_3} stress and performing a sufficient number of iterative experiments (indoor), and estimating complex influence from different environmental stressors (outdoor). The lower range of generated ozone in the \ce{O_3}-air mix is about 30 \ce{\mu gm^{-3}} over the atmospheric level, where all plants demonstrated a stable response. Measurement results indicate a well detectable reaction of hydrodynamic system to changes of \ce{O_3} concentration in the upper part of stem (measured by upper EIS sensor) with a delay of 10-20 minutes between the onset of exposure and biological response. The difference between low- and high-ozone days is also detectable outdoors with 40-50 \ce{\mu gm^{-3}} difference between low/high \ce{O_3} cases. By combining data from different plants in both setups, the detection rate was increased to 92\% and low/high \ce{O_3} cases were clearly separated for pattern recognition algorithms and machine learning techniques. There are several methodological issues related to recognition of physiological reactions, \ce{O_3} generation and protection of plants in outdoor setup -- they can be improved in further attempts. EIS sensors prove to be extremely reliable, especially in harsh outdoor conditions. They also seem to be well suited for environmental monitoring and biosensing in addition to applications in precision agriculture and vertical farming.

\section{Acknowledgement}

This research was funded by European Union under Horizon 2020 research and innovation program, Grant Agreement No. 101017899, ‘WATCHPLANT: Smart Biohybrid Phyto-Organisms for Environmental In Situ Monitoring’. Author would like to thank Antonio Diaz Espejo (Institute for Natural Resources and Agrobiology, IRNAS-CSIC, Sevilla, Spain) for providing tobacco seeds and fruitful discussions about plant physiology as well as all members of CYBRES team and WATCHPLANT consortium for technical discussions on phytosensing and setup development.

\section{Conflicts of Interest}

The author declares that the research was conducted in the absence of any commercial or financial relationships that could be construed as a potential conflict of interest. The funders had no role in the design of the study; in the collection, analyses, or interpretation of data; in the writing of the manuscript; or in the decision to publish the results.

\small

\newpage
\appendix

\begin{figure*}[htp]
\centering
\subfigure[]{\includegraphics[width=.49\textwidth]{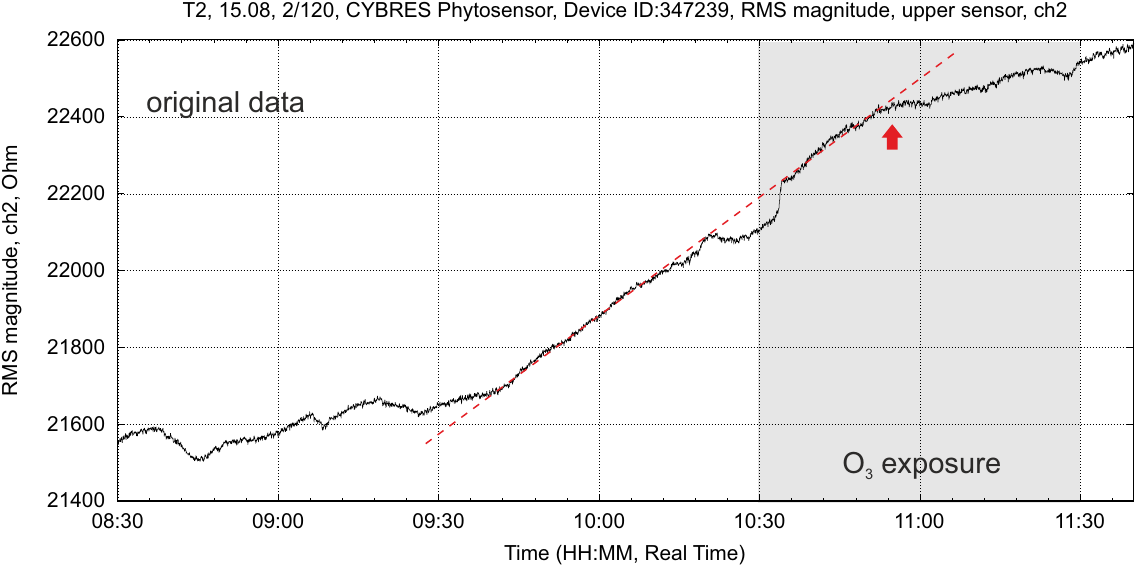}}~
\subfigure[]{\includegraphics[width=.49\textwidth]{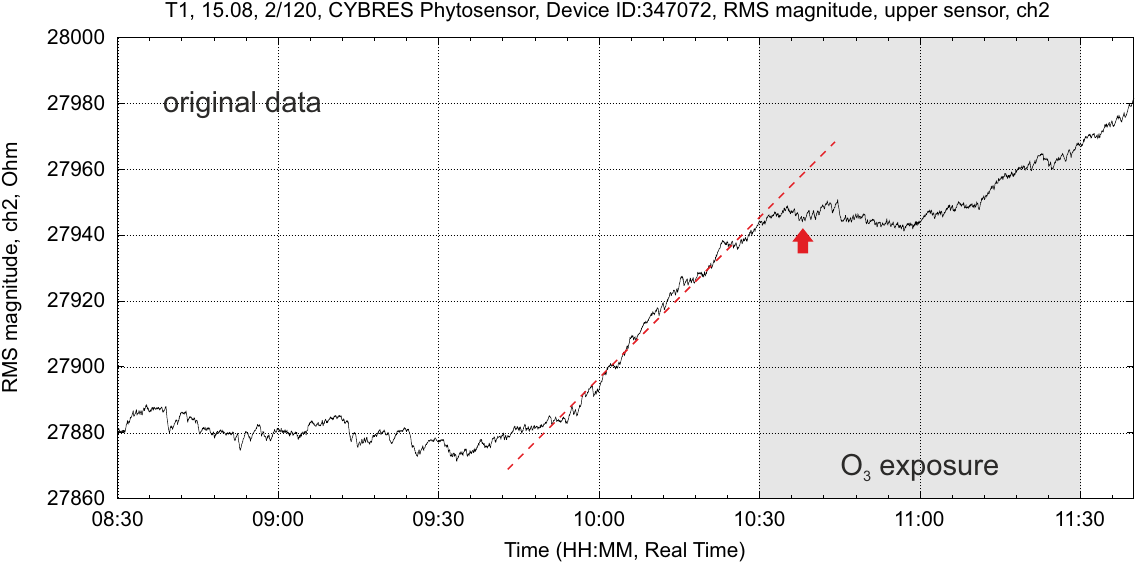}}
\subfigure[]{\includegraphics[width=.49\textwidth]{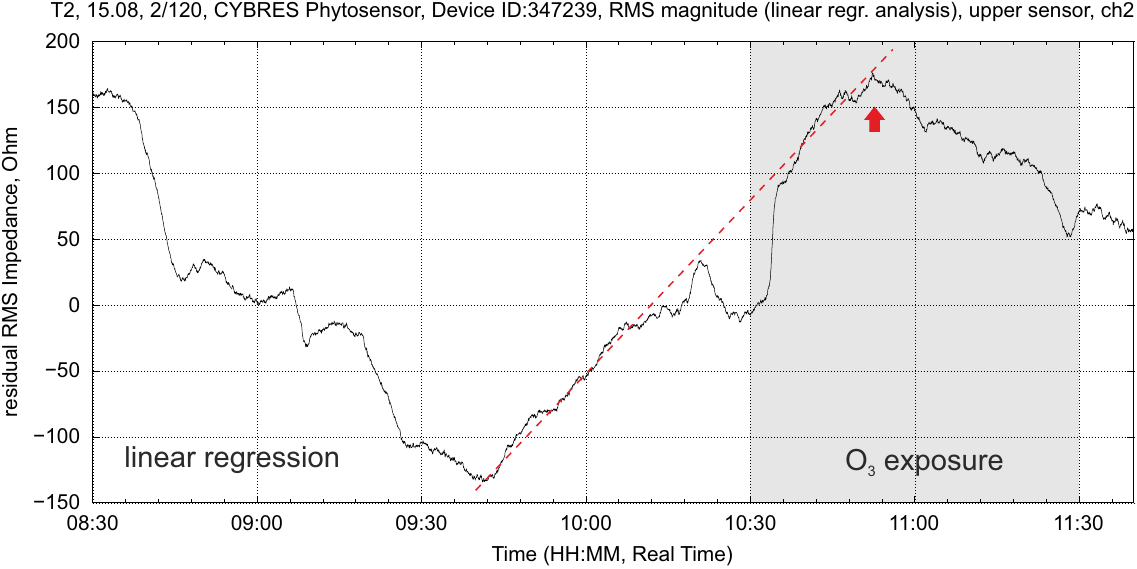}}~
\subfigure[]{\includegraphics[width=.49\textwidth]{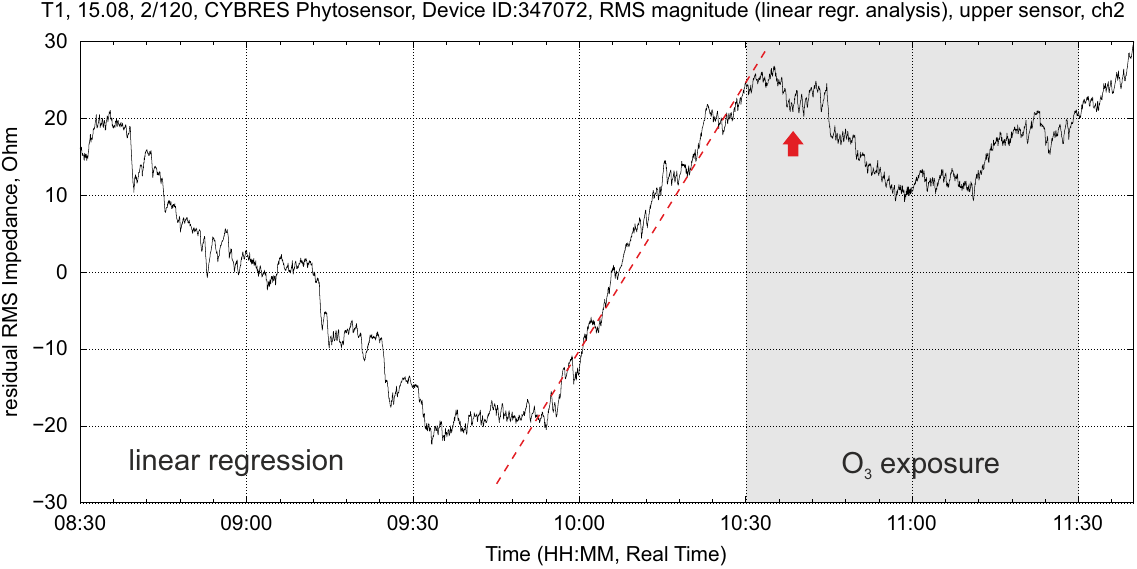}}
\subfigure[]{\includegraphics[width=.49\textwidth]{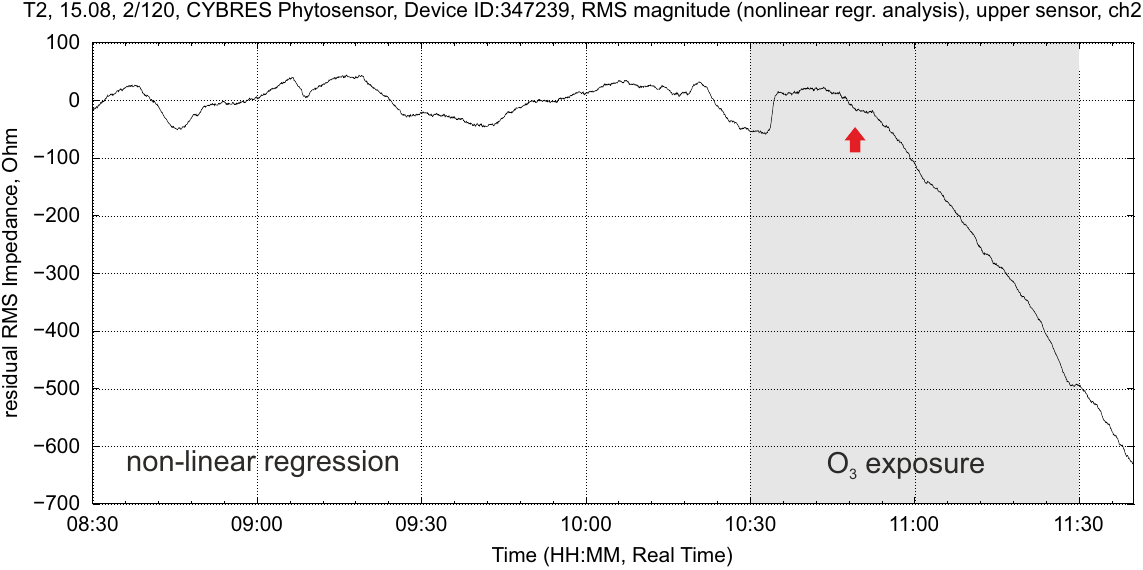}}~
\subfigure[]{\includegraphics[width=.49\textwidth]{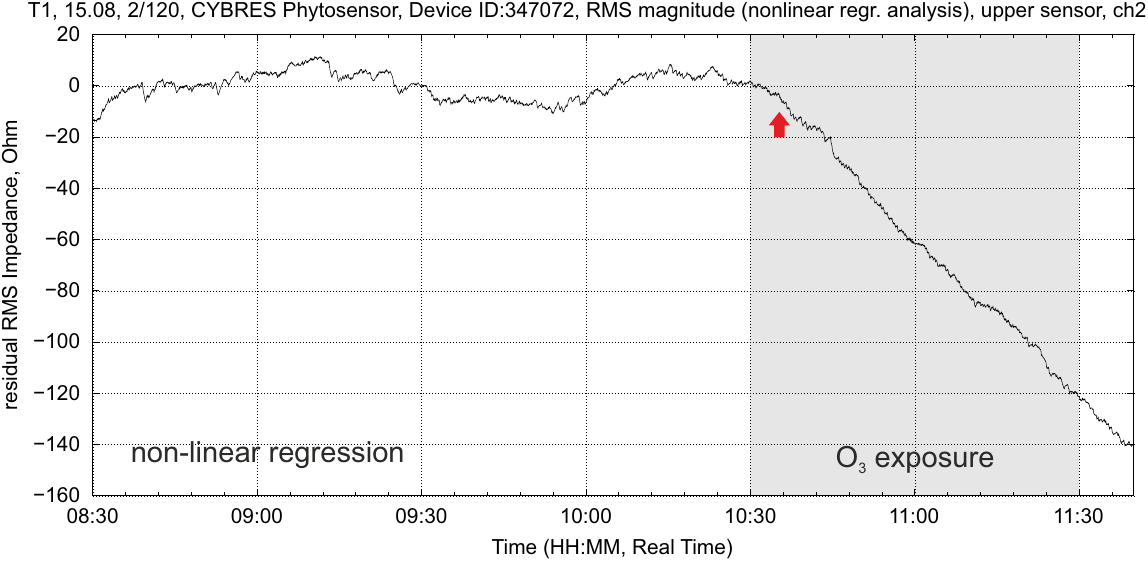}}
\caption{\small Supplementary image: demonstration of regression analysis for detecting changes of the EIS trend, tobacco T1 and T2 plants, \ce{O_3} exposure in 15 ppb range: \textbf{(a,b)} Original EIS dynamics of upper sensors; \textbf{(c,d)} linear regression up to the inclination point (interpolation after this point); \textbf{(e,f)} nonlinear regression with 5th order polynome up to the inclination point (interpolation after this point). In all cases the inclination point is detectable, nonlinear regression provides the best resolution for detection based on $\Psi<0$ condition, see the expression (\ref{eq:cond1}).  
\label{fig:exp_1508}}
\end{figure*}	

\begin{figure*}[htp]
\centering
\subfigure[]{\includegraphics[width=.49\textwidth]{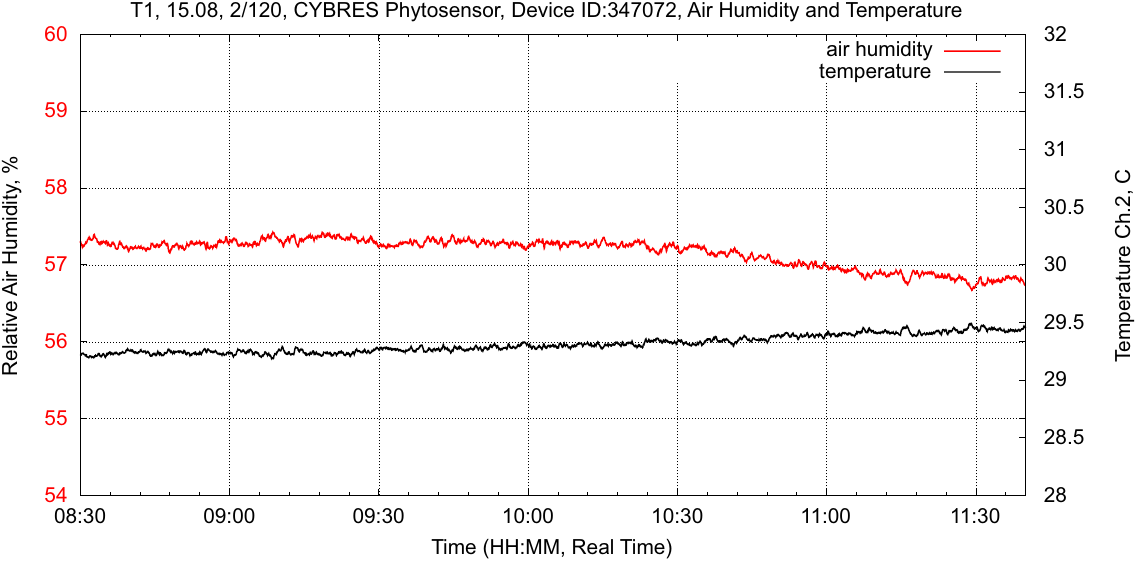}}~
\subfigure[]{\includegraphics[width=.49\textwidth]{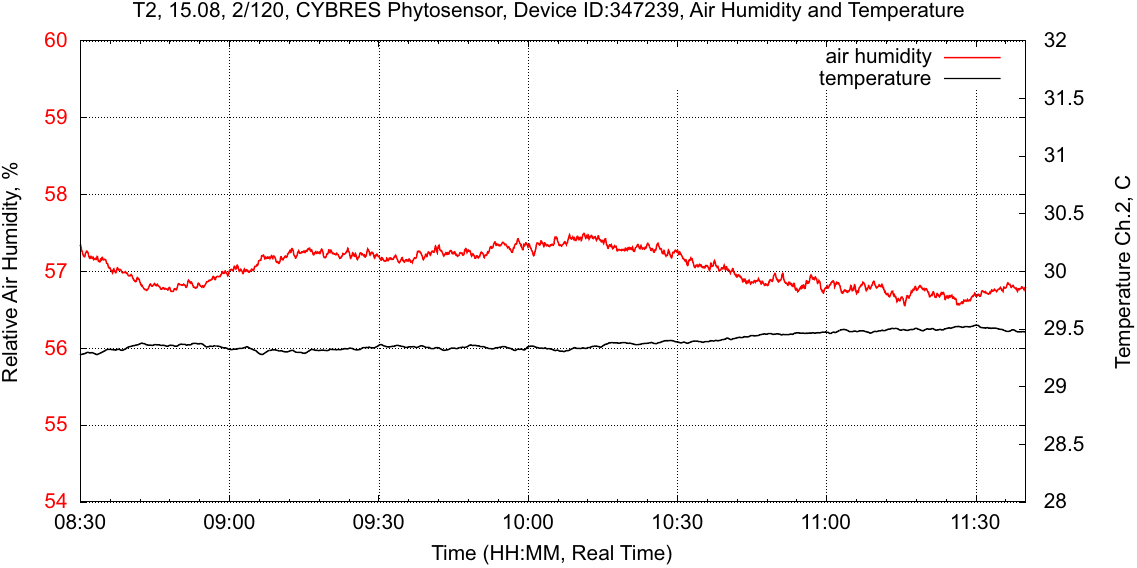}}
\subfigure[]{\includegraphics[width=.49\textwidth]{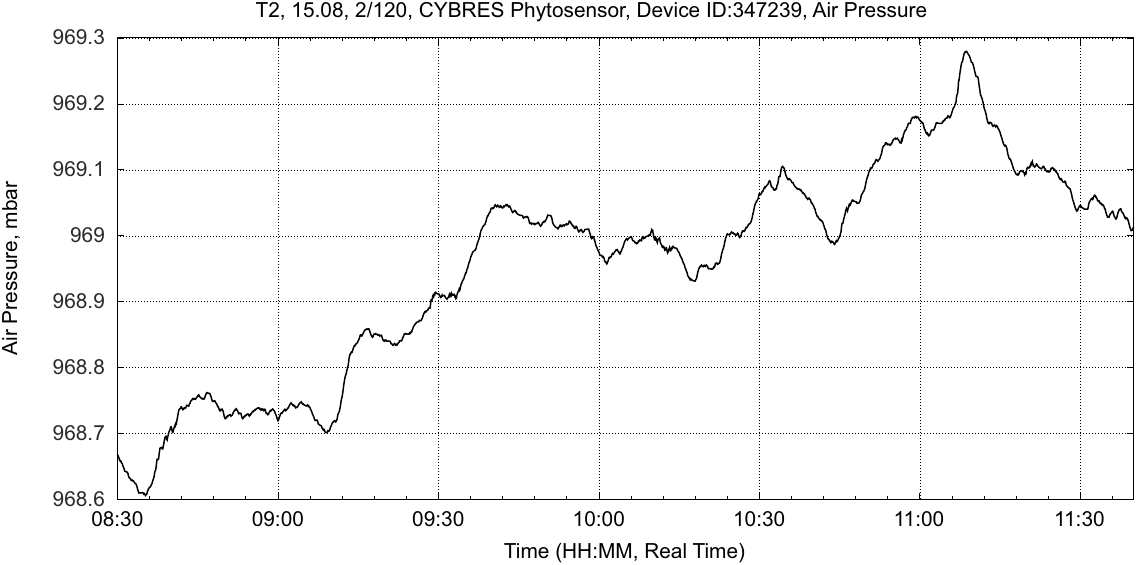}}~
\subfigure[]{\includegraphics[width=.49\textwidth]{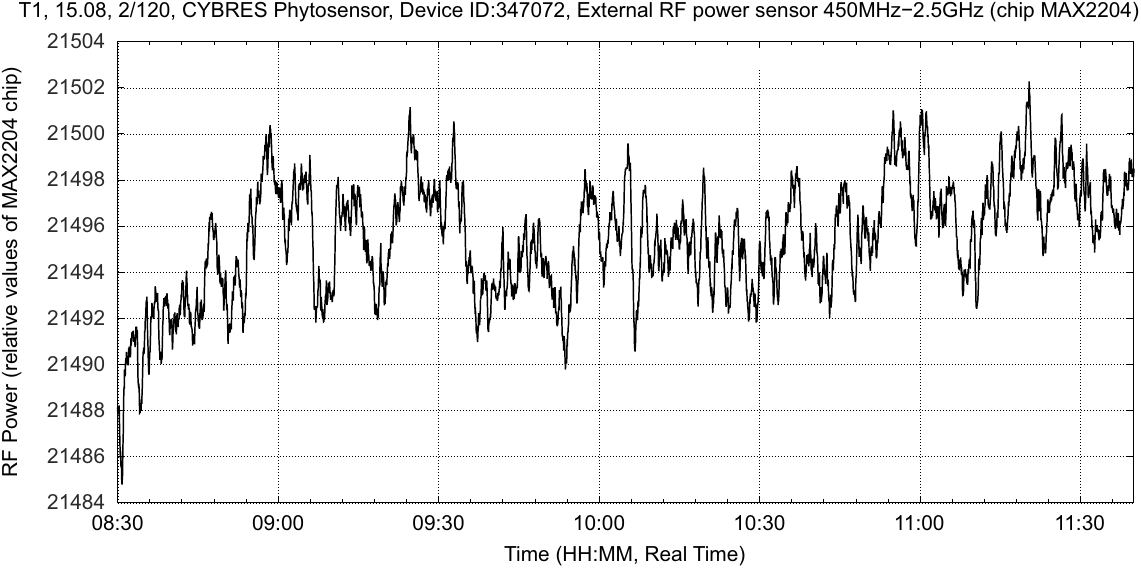}}
\subfigure[]{\includegraphics[width=.49\textwidth]{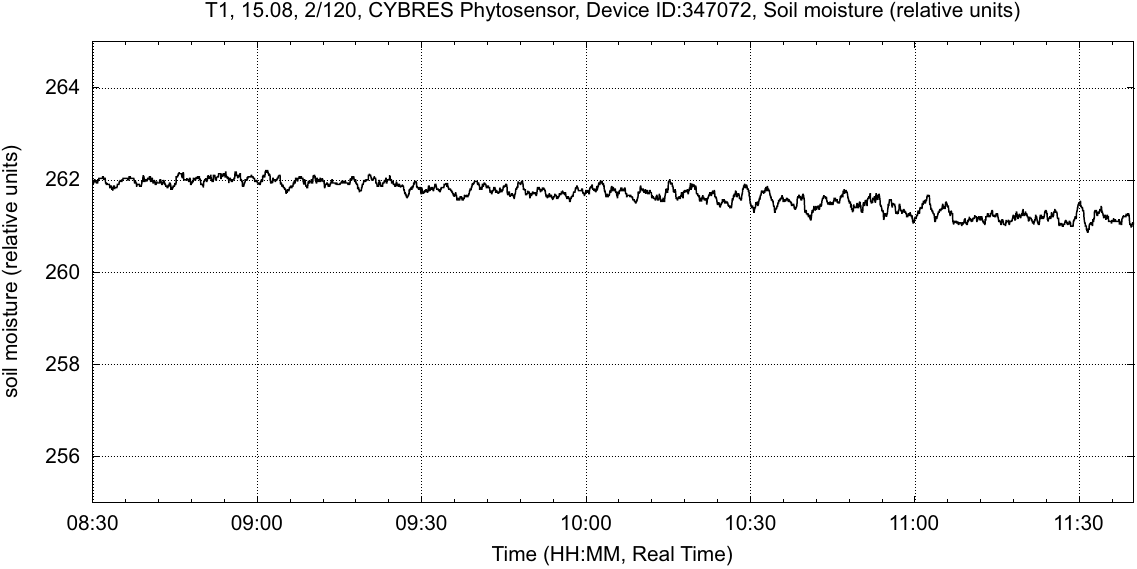}}
\caption{\small Supplementary image: environmental parameters (air humidity and temperature, atmospheric pressure, EM emission in 450Mhz-2.5Ghz range, soil moisture) for the EIS measurements in Fig. \ref{fig:exp_1508} before and during \ce{O_3} exposure. \label{fig:exp_1508_add}}
\end{figure*}	

\begin{figure}[htp]
\centering
\subfigure{\includegraphics[width=.49\textwidth]{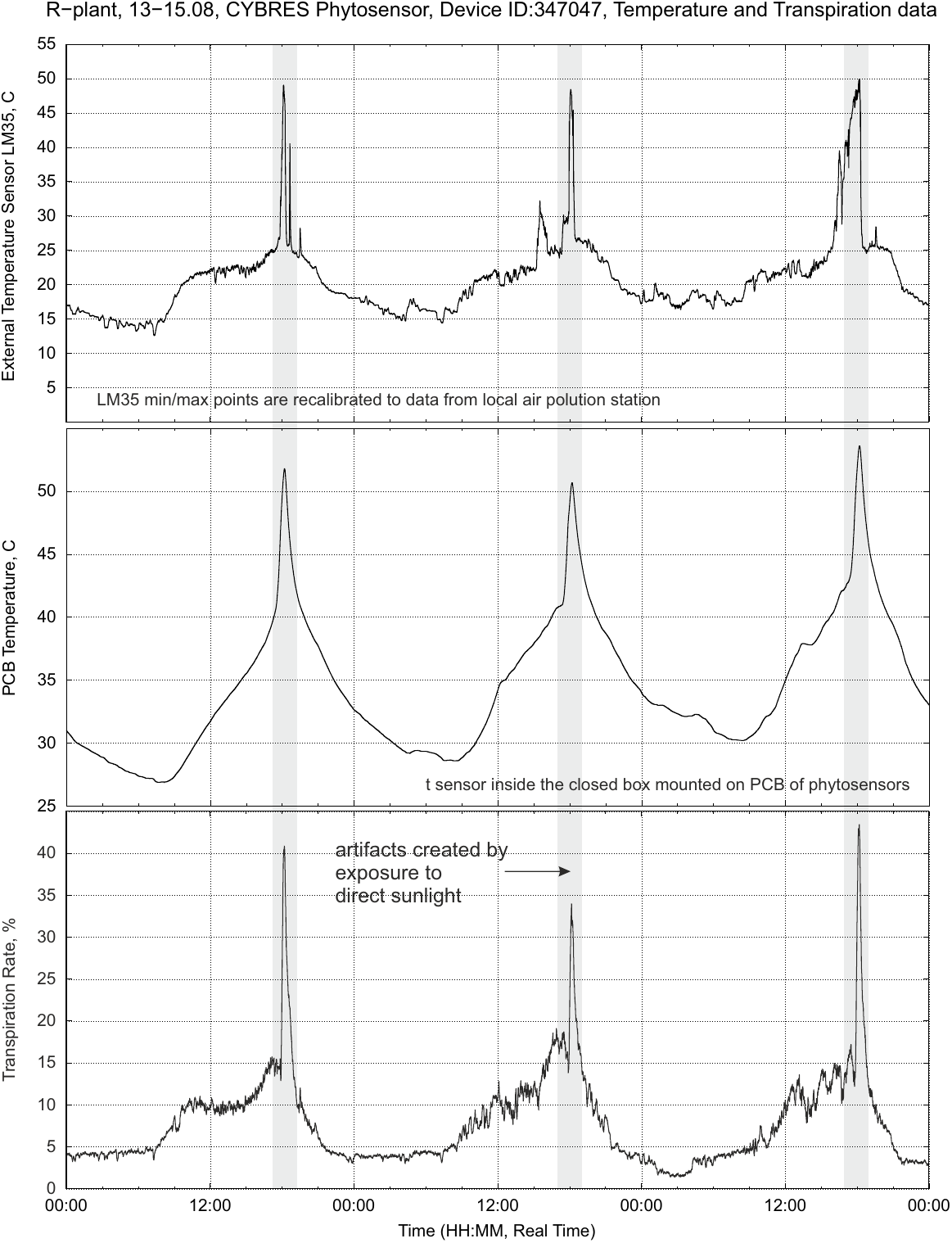}}
\caption{\small Supplementary image for Fig. \ref{fig:heatStress}: environmental temperature (rescaled to data from local air pollution stations), internal temperature of electronic components (inside the closed box) and transpiration in outdoor setup during 72 hours, shadow regions represent a direct sunlight exposure in the afternoon time. \label{fig:heatStressAdd}}
\end{figure}

\begin{figure}[htp]
\centering
\subfigure{\includegraphics[width=.49\textwidth]{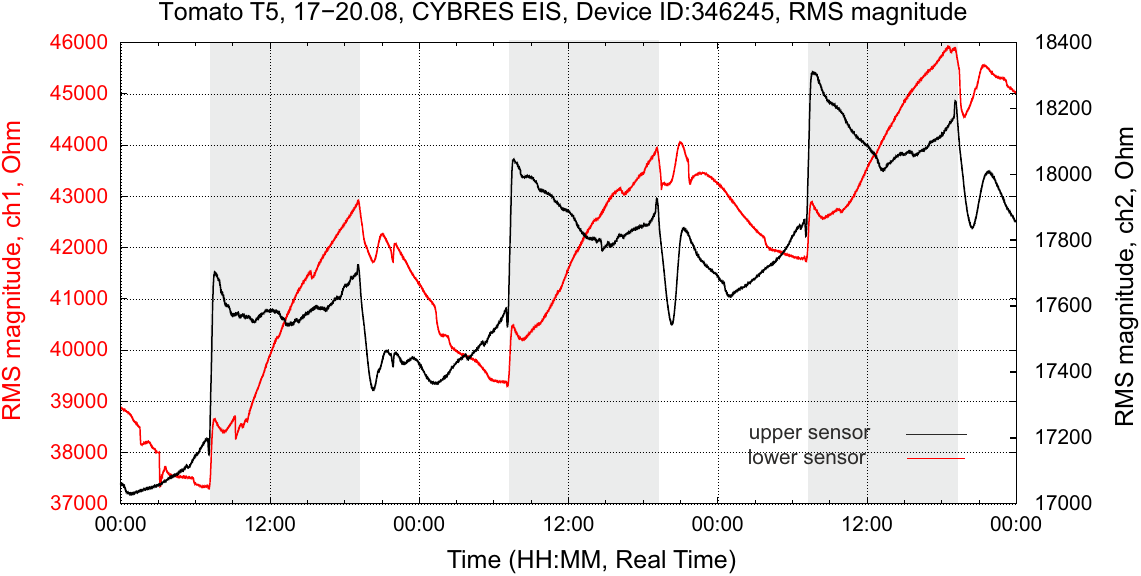}}
\caption{\small Supplementary image: EIS dynamics of tomato plant (indoor setup) during 72 hours. \label{fig:tomatoIndoor}}
\end{figure}

\begin{figure}[htp]
\centering
\subfigure{\includegraphics[width=.49\textwidth]{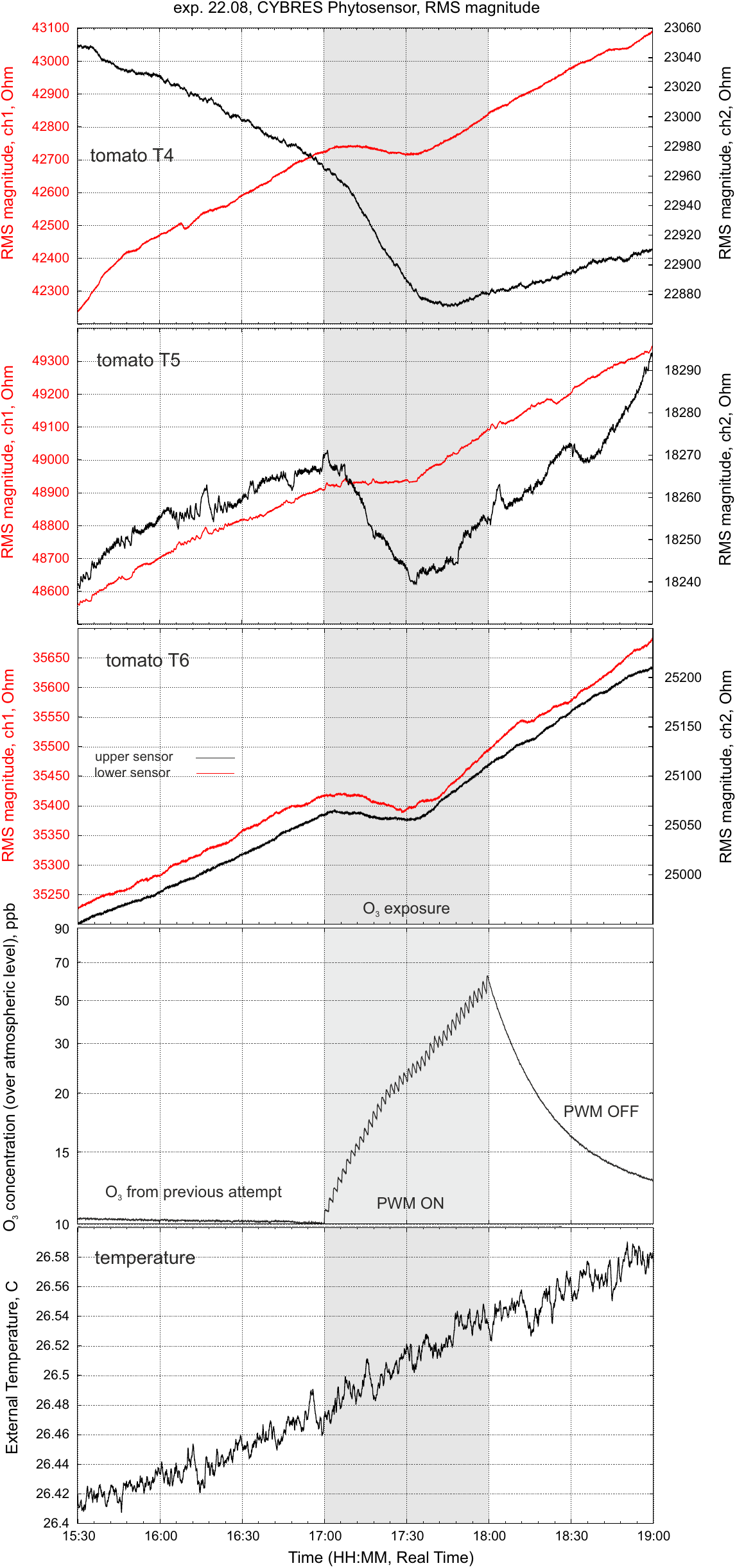}}
\caption{\small Reaction of three different tomato plants on \ce{O_3} exposure (indoor setup), variation of temperature is about 0.2C during 3.5 hours.
\label{fig:homogeneousTomato}}
\end{figure}

\begin{figure}[htp]
\centering
\subfigure{\includegraphics[width=.49\textwidth]{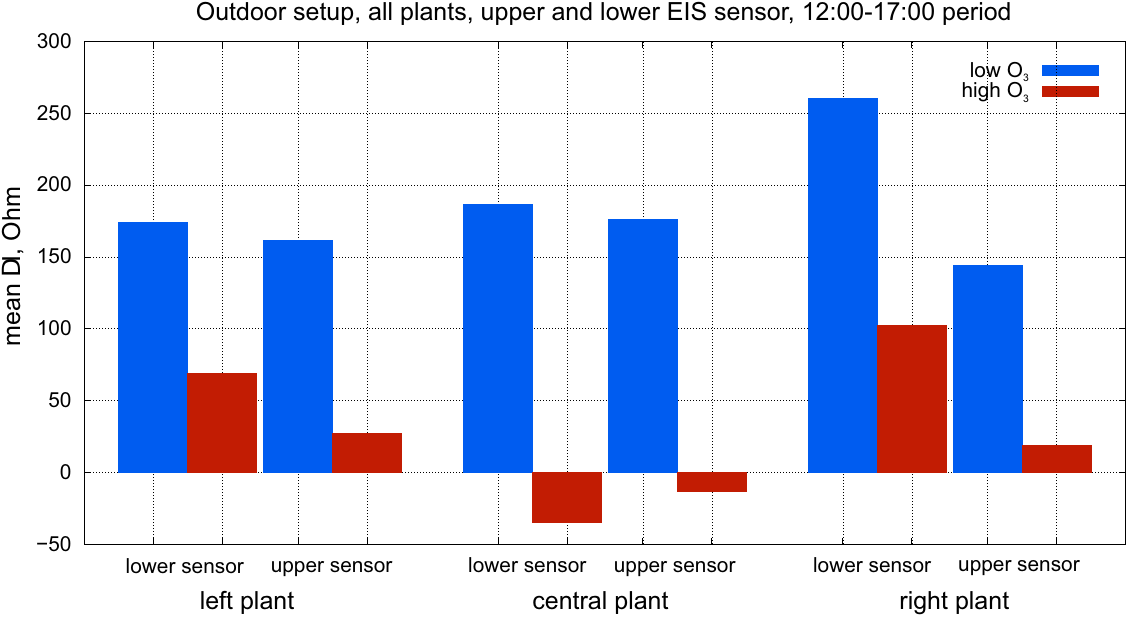}}
\subfigure{\includegraphics[width=.49\textwidth]{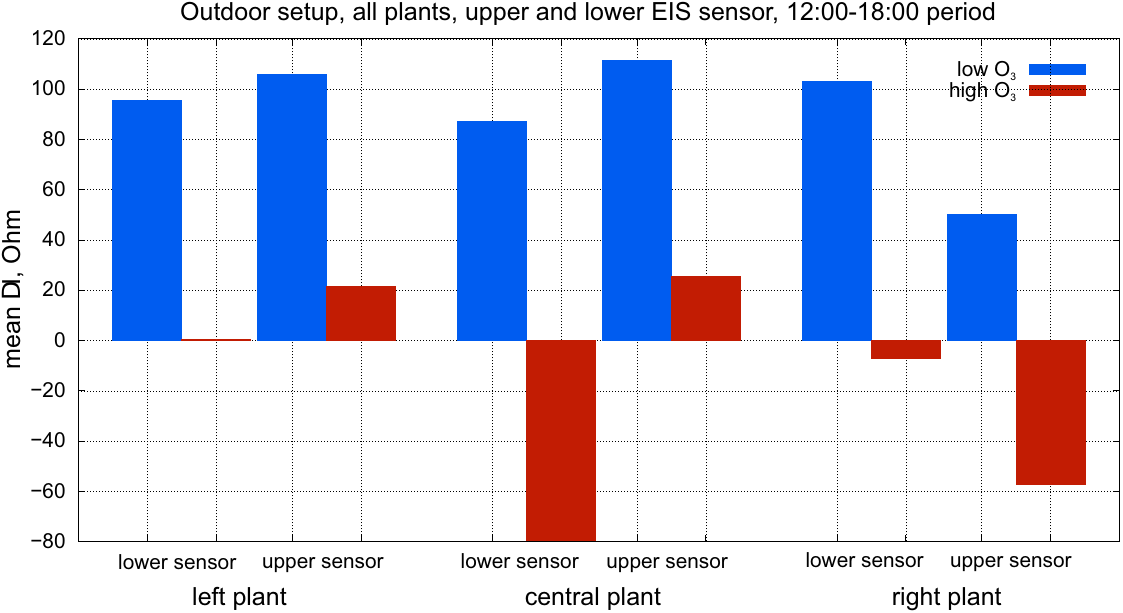}}
\caption{\small Comparison of means for all sensors and plants for 12:00-17:00 and 12:00-18:00 measurement regions (with heat stress). \label{fig:outdoor_stats3}}
\end{figure}

\end{document}